\documentclass[aps,pre,twocolumn,amsmath,floatfix,showpacs]{revtex4}

\usepackage{graphicx}
\usepackage[usenames]{color}
\usepackage{soul}

\begin{document}
\title{Relaxation dynamics in a transient network fluid with competing gel and glass phases}
\author{Pinaki Chaudhuri} 
\affiliation{The Institute of Mathematical Sciences, C.I.T. Campus, Taramani, Chennai 600 113, India}
\author{Pablo I. Hurtado}
\affiliation{Instituto Carlos I de F\'{\i}sica Te\'orica y Computacional,
and Departamento de Electromagnetismo y F\'{\i}sica de la Materia,
Universidad de Granada, Granada 18071, Spain}
\author{Ludovic Berthier}
\affiliation{Laboratoire Charles Coulomb, UMR~5221, Universit\'e Montpellier and CNRS, 34095 Montpellier, France}
\author{Walter Kob}
\affiliation{Laboratoire Charles Coulomb, UMR~5221, Universit\'e Montpellier and CNRS, 34095 Montpellier, France}

\begin{abstract}
We use computer simulations to study the relaxation dynamics of a
model for oil-in-water microemulsion droplets linked with telechelic
polymers. This system exhibits both gel and glass phases and we show
that the competition between these two arrest mechanisms can result in a
complex, three-step decay of the time correlation functions, controlled
by two different localization lengthscales. For certain combinations of
the parameters, this competition gives rise to an anomalous logarithmic
decay of the correlation functions and a subdiffusive particle motion, which
can be understood as a simple crossover effect between the two relaxation
processes. We establish a simple criterion for this logarithmic decay
to be observed. We also find a further logarithmically slow relaxation
related to the relaxation of floppy clusters of particles in a crowded
environment, in agreement with recent findings in other models for dense
chemical gels. Finally, we characterize how the competition of gel and
glass arrest mechanisms affects the dynamical heterogeneities and show
that for certain combination of parameters these heterogeneities can be
unusually large. By measuring the four-point dynamical susceptibility,
we probe the cooperativity of the motion and find that
with increasing coupling this cooperativity shows a maximum before it
decreases again, indicating the change in the nature of the relaxation
dynamics. Our results suggest that compressing gels to large densities
produces novel arrested phases that have a new and complex dynamics.
\end{abstract}

\maketitle

\section{Introduction}

In nature and in our daily life, many soft materials are
formed due to the dynamical arrest of the constituent particles
\cite{book,witten,tartaglia}. Usually they are labelled as gels if
the particle density is low and as glasses if the density is large.
However, the difference between these two states is at present not very
well understood and therefore it is not always easy to distinguish them.
Despite this difficulty quite a few features in this glass-gel cross-cover
regime have been studied extensively.

In dense glass forming liquids the slowing down of dynamics is related to
the mutual steric-hindrance in the motion of the constituent particles.
The dynamical properties of these glass formers are characterized by
a stretched-exponential shape of relaxation functions \cite{book},
or similarly by the anomalous, exponential tails in the van Hove
distributions of particle displacements \cite{Pinaki}. Many of
these dynamical features are described well by mode-coupling theory
\cite{book}. At even lower temperatures, the relaxation dynamics can
be understood by means of the random first order transition theory
\cite{rfot}. In these glass formers, the structure is usually close-packed
for hard sphere or van der Waals type interactions and is accompanied
by a super-Arrhenius increase of viscosity. Or, if there are covalent
bondings, they form network-like structures and exhibit an Arrhenius
increase in viscosity.

On the other hand, chemical gels or rubbers are soft solids having random
network structures \cite{witten,larson,zaccarellirev}. The cross linking
of the permanent bonds between the constituent monomers happens during the
synthesis process inducing a vulcanization transition once the density of
the links exceeds the percolation threshold~\cite{zippeliusrev}. Different
static and dynamical properties in the vicinity of this transition
have been studied, both using simulations and theoretical models (e.g.,
see \cite{coniglio1, zippelius1}). Physical gels are on the other hand
low-density network structures with bonds that can be broken/realigned by
thermal fluctuations within finite timescales~\cite{zaccarellirev}. One
possible nonequilibrium path to physical gelation is via a thermal
quench across the liquid-gas spinodal leading to dynamically
arrested states \cite{zaccarelli2, paddy0, paddy1,paddy2}, that
show complex aging phenomena \cite{testard,suarez}.  In general,
these paths lead to spatially heterogeneous structures. However,
in recent times, considerable effort has been made to devise
ways by which spatially homogeneous physical gels can be formed
\cite{pablo,porte,pablo2,pre2010,shibu,delgado1,coniglio1,coniglio2,miller,patch1,patch2}.
For such gel-forming systems, a wide variety of relaxation functions have
been reported: logarithmic \cite{zaccarelli3, zaccarelli4, puertas,log},
stretched \cite{suarez} or compressed exponentials \cite{duri, shibu}.
While theoretical models \cite{fuchs, dawson, kroy, patch2} have been
proposed to account for such dynamical properties, they are at this time
certainly not yet comprehensive.

Of particular interest is the interplay between these different arrest
mechanisms, viz. gel and glass, since their competition can be used
to engineer materials with novel functionalities \cite{pre2010, zhao, hoffmann, royalltanaka}. 
Similar studies have been
carried out in systems with competing lengthscales \cite{moreno,sperl}
or interactions \cite{reviewfuchs,moreno-polymer,zaccarelli3}. Yet,
only few models do allow to study the low density gel phase and high density
glass at the same time. Accessing this regime is, however, necessary for
investigating the structure and dynamics at those intermediate densities
where the gel transforms into a glass and vice versa. Here, we study
a simple model with direct experimental relevance \cite{porte}, and
which permits us to traverse the density regime of interest and hence
to study the interplay of different processes which lead to either
gelation or glassiness.  Furthermore, our model also allows us to tune
the lifetime of bonds, a feature that is usually not present in other
models (for example, see the recent work \cite{Nagi}). On one hand,
this facilitates a wider exploration of the relaxation dynamics of such
model physical gels, but also allows to disentangle the origin of the
apparently anomalous relaxation observed in these systems. In fact,
mode-coupling theory predicts, e.g., logarithmic relaxation whenever
two different arrest lines meet, one gel-like and another glass-like,
and relates it to an underlying higher-order singularity in the theory
\cite{dawson}.  The versatility of our model, and in particular the
possibility of tuning at will the bonds lifetime, allows us to explore
the different mechanisms at play and hence to elucidate the interplay
of various lengths and time-scales \cite{pre2010}.

The paper is structured as follows. In Section II we explain the
details of our model transient-network fluid, together with the
numerical schemes used to simulate its dynamics.  The phase diagram
of the model system and its structural properties are discussed in
Sections III and IV, respectively. In Section V, we analyze in
details the fluid's dynamics, quantified by the mean squared
displacement and the incoherent scattering function.  For that, we
follow different routes across the phase diagram, which allows us
to clearly understand the interplay between the gel and the glass
regimes. Finally, the dynamical heterogeneities which characterize
the slow dynamics in the gel and glass phases are studied in Section
VI, followed by a summary of our results and a broader perspective
in Section VII.

\section{Model and details of simulation}

Our model system is a coarse-grained representation~\cite{pablo} of a
transient gel which has been studied in experiments~\cite{porte}. In this
system, an equilibrium low-density gel is obtained by adding telechelic
polymers to an oil-in-water microemulsion. Since the polymer end-groups
are hydrophobic, the polymers effectively act as (attractive) bridges
between the oil droplets they connect. The strength, lengthscale and
typical lifetime of these bridging polymers can be controlled at will.
Denoting by $C_{ij}$ the number of polymers connecting droplets $i$
and $j$, we have established in Refs.~\cite{pablo,pablo2}, that the
following interaction is a reasonable coarse-grained representation of
this ternary system:

\begin{equation}
V =  \epsilon_1 \sum_{j > i}
\left( \frac{\sigma_{ij}}{r_{ij}} \right)^{14}
+ \epsilon_2 \sum_{j > i}
C_{ij} V_{\rm FENE}(r_{ij})
+\epsilon_{0} \sum_{i} C_{ii}.
\label{model}
\end{equation}

\noindent

The first term is a soft repulsion acting between bare oil droplets, where
$\sigma_{ij} = (\sigma_i + \sigma_j)/2$, $\sigma_i$ is the diameter of
droplet $i$, and $r_{ij}$ is the distance between the droplet centers.
The second term describes the entropic attraction induced by the
telechelic polymers, which has the standard ``FENE'' (finitely-extensible
nonlinear elastic) form known from polymer physics~\cite{witten},
$V_{\rm FENE}(r_{ij}) = \ln ( 1 - (r_{ij}-\sigma_{ij})^2/\ell^2)$,
and accounts for the maximal extension $\ell$ of the polymers. The
last term introduces the energy penalty $\epsilon_0$ for polymers
that have both end-groups in the same droplet.  The most drastic
approximation of the model (\ref{model}) is the description of the
polymers as effective bonds between the droplets, which is justified
whenever the typical lifetime of the bonds is much larger than the
timescale for polymer dynamics in the solvent~\cite{porte}.  Thus,
for exploring the different properties of such a model, the relevant
variables are the droplet volume fraction $\phi$ and the number of
polymer heads per droplet $R$ \cite{pablo}.  In order to describe the
dynamics of the system, we use a combination of molecular dynamics to
propagate the droplets with the interaction (\ref{model}), and local
Monte Carlo moves with Metropolis acceptance rates $\tau_{\rm link}^{-1}
{\rm min}[1,\exp(\Delta V/k_BT)]$ to update the polymer connectivity
matrix $C_{ij}$, where $\Delta V$ is the difference in potential energy
of the system for the two bond configurations~\cite{pablo,pablo2}. Thus
$\tau_{\rm link}$ is the timescale governing the renewal of the polymer
network topology.  In order to prevent crystallization at high volume
fractions (which would be the case for the monodisperse model studied
earlier \cite{pablo}), we use a polydisperse emulsion with a flat
distribution of particle sizes in the range $\sigma_i \in [0.75, 1.25]$
(having a mean diameter $\sigma=1$).  The units of length, energy and time
are respectively $\sigma$, $\epsilon_1$ and $\sigma\sqrt{m/\epsilon_1}$
where $m$ is the mass of the particles. The space of control parameters
is quite large.  Therefore we set $\ell=3.5 \, \sigma$ as measured in
experiments~\cite{porte}, $T=1$, and $\epsilon_0=1$ and $\epsilon_2=50$,
and vary the remaining parameters $\{\phi, R, \tau_{\rm link}\}$.
These choices for the parameter values 
leads to a phase diagram which is similar
to the one obtained in experiments.

Our numerical simulations are done for a three dimensional system
of $N=1000$ particles. The equations of motions of these particles
are integrated using a velocity Verlet scheme with a time step of
$\delta{t}=0.005$.  Here, most of the results are reported for
$\tau_{\rm link}=10^2$, although we also explore other values of
$\tau_{\rm link}$: $1,10,10^3,10^4$ to illustrate some of the dynamical
features of the system.  At each volume fraction $\phi$, we first
equilibrate the system of particles without any links ($R=0$). Once
equilibrated, bonds are introduced corresponding to the required
value of $R$ and then the system is again equilibrated to obtain
the proper distribution of bonds per particle.  Since the structure
of the network is independent of the choice of $\tau_{\rm link}$
\cite{pablo2}, we use a small value of $\tau_{\rm link}=1$ to expedite the
equilibration process.  Subsequently, data is generated by continuing the
simulations with different $\tau_{\rm link}$ values when required
and the averages are typically calculated over 100 different time origins.
We also do simulations for the
case when the bonds between two particles are completely frozen. In order
to do a proper sampling of the network configurations for this situation,
we use 6 initial configurations (positions,
connectivities) from the simulations with a finite $\tau_{\rm link}$
as initial inputs for subsequent evolution of the particle positions
using molecular dynamics with the connections now permanently fixed.

\section{Phase diagram}

\begin{figure}[]
\includegraphics[width=80mm,clip=true]{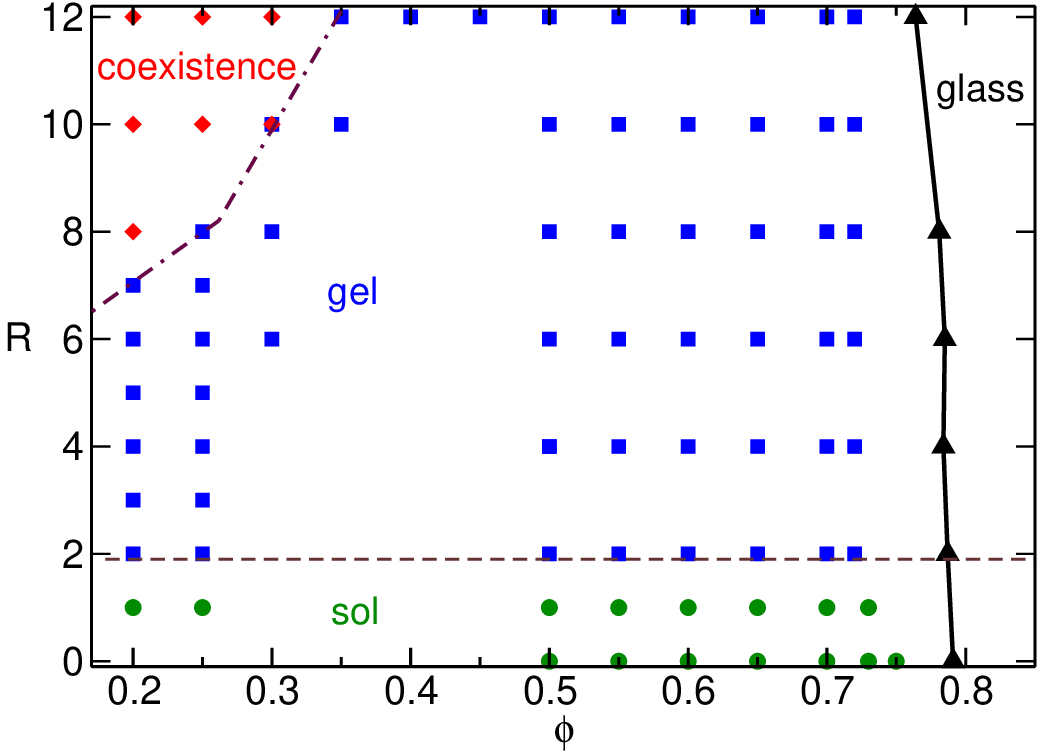}
\caption{(Color online) 
Phase diagram of the system obtained by varying $R$ and
$\phi$. The diamonds correspond to the coexistence region between gas
and liquid, the circles to the sol phase and the squares to the gel phase. 
The horizontal dashed line corresponds to onset of percolation, the dot-dashed line
indicates the phase-coexistence boundary and the thick line
marks the predicted glass line. Snapshots of typical configurations in 
each phase have been published elsewhere, see Fig. 3 in \cite{pablo2}.}
\label{phasediagram}
\end{figure}

We begin by summarizing our earlier findings for the phase diagram
(shown in Fig.~\ref{phasediagram}) for this model. If the number of
polymers is small (i.e $R<2$), the system is in a simple liquid phase
(the sol) at small values of $\phi$. In this regime, the distribution
of connectivities per particle is just an exponential \cite{pablo2}.
With increasing $\phi$, the dynamics in the sol regime becomes slow and
one eventually enters a glassy phase at large $\phi$, characterized by
a very strong increase of the timescales for structural relaxation. If
the number of bonds is large and $\phi$ is small, phase separation
is observed due to strong attractions between the droplets. Here, the
distribution of connectivities becomes bimodal, with one of the two peaks
corresponding to a well-connected liquid and the other to free particles.
For intermediate values of $R$ the system is in a gel phase.  In this
region of the phase diagram the particles are connected together and
form a percolating cluster and the spatial density is homogeneous. Here,
the connectivity distribution is peaked around the average value and has
an exponential tail. Note that this gel is an equilibrium phase since the
polymer network is constantly rewired on the timescale $\tau_{\rm link}$.
However, if we go to large enough volume fractions, we observe again a
glassy system for all connectivities, with the corresponding divergence
of relaxation timescales.

At low $\phi$, in the gel phase, the main slow relaxation process is
related to the connections between the droplets by means of the polymers
and the timescale associated with its reconfiguration. At low $R$, as the
system becomes glassy at large $\phi$, the origin of the slow dynamics
is the steric hindrance caused by the caging of each particle by its
neighbors. In the region where both $R$ and $\phi$ act as a source for
slow dynamics, we have shown that the generic relaxation process has three
steps but with proper tuning of the two relaxation timescales, one can
also obtain logarithmic decays of the relaxation function~\cite{pre2010}.

In the following we will discuss in details the interplay between these
two relaxation processes and show the consequences on the nature of
different dynamical quantities as we move around in the phase diagram.

\section{Structure}

\begin{figure}[h]
\includegraphics[width=80mm,clip=true]{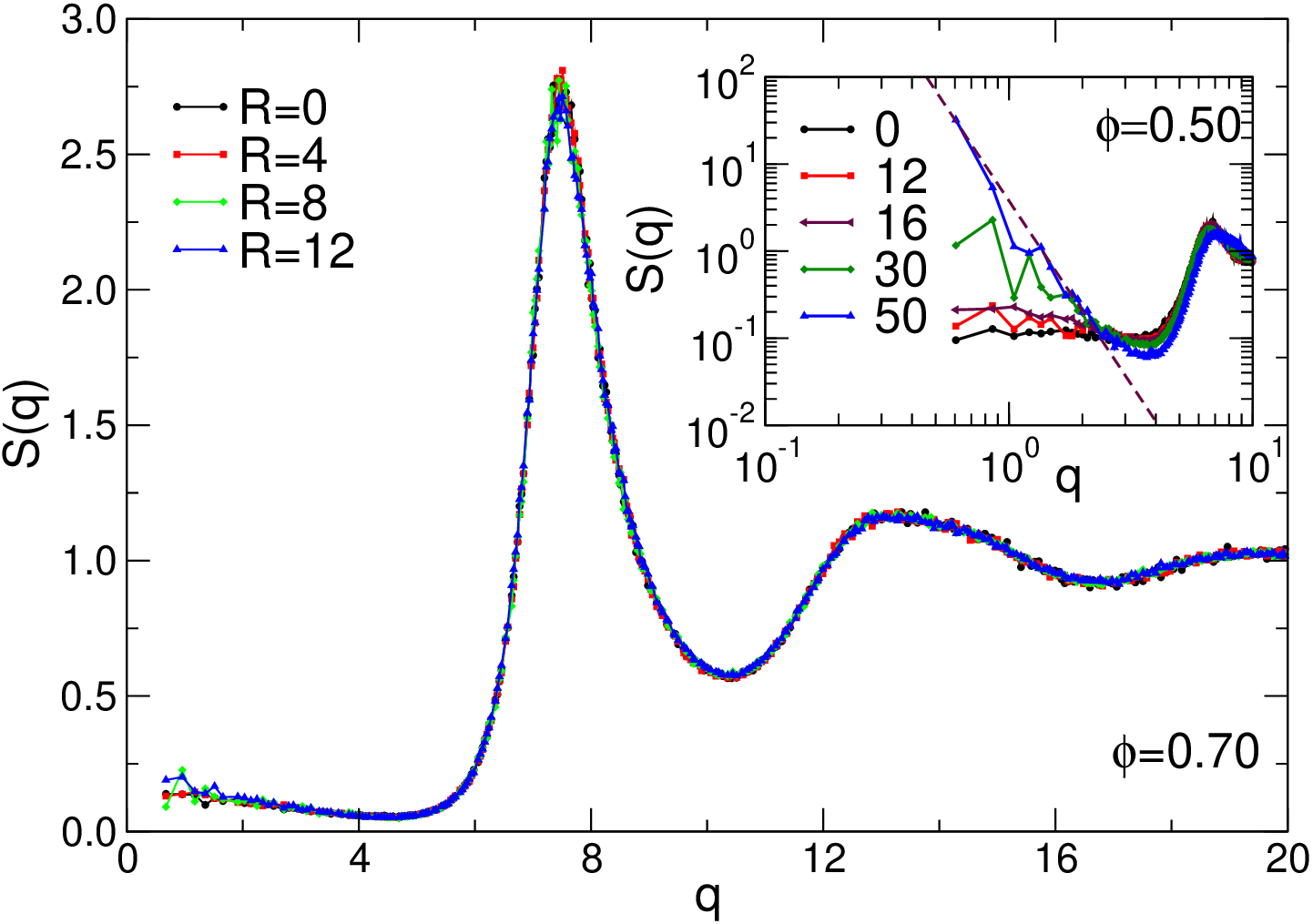}
\caption{(Color online) Main panel: Structure factor $S(q)$ computed for the particles at a 
volume fraction of $\phi=0.70$ for different connectivities $R=0,4,8,$ and 12.
Inset: $S(q)$ at $\phi=0.50$ for $R=0,12,30,$ and 50. The dashed line corresponds to $q^{-4}$.
} 
\label{sq70}
\end{figure}

Before we discus the different dynamical properties of the system, we
briefly look at its structure.  In Fig.~\ref{sq70} we plot the static
structure factor, $S(q)$, for a system that is dense, $\phi=0.70$,
varying the connectivity $R$. The general shape of $S(q)$ is very similar
to the one of a simple liquid and hence we can conclude that the system
is homogeneous. Also, we see no significant dependence of $S(q)$ on $R$,
thus showing that at high densities the structure is mainly governed by
steric hindrance. For intermediate densities, however, we do note the
weak dependence on $R$ in the regime of small wave-vectors, illustrated
in the inset of Fig.~\ref{sq70} where we show $S(q)$ for $\phi=0.50$.
This is at a volume fraction at which for large $R$ the system approaches
the co-existence region, and hence one starts to see the emergence of
a power-law behavior at small values of $q$ with increasing $R$; the
data for $R=50$ can be approximated by $q^{-4}$ (which is expected in
proximity to phase co-existence \cite{furukawa}).

\section{Relaxation dynamics}

\subsection{Dependence on volume fraction $\phi$}

\begin{figure}[t]
\includegraphics[width=80mm,clip=true]{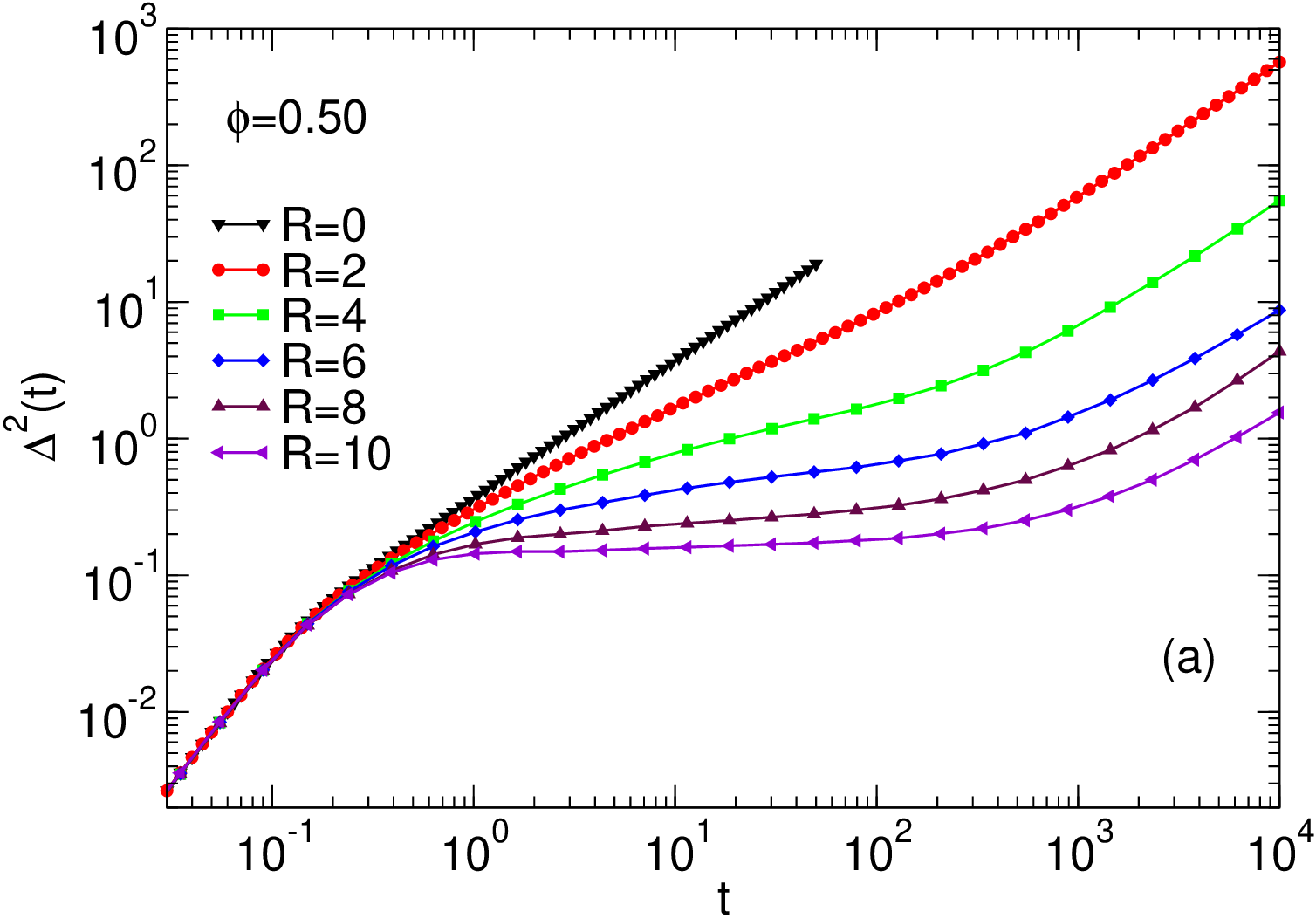}
\includegraphics[width=80mm,clip=true]{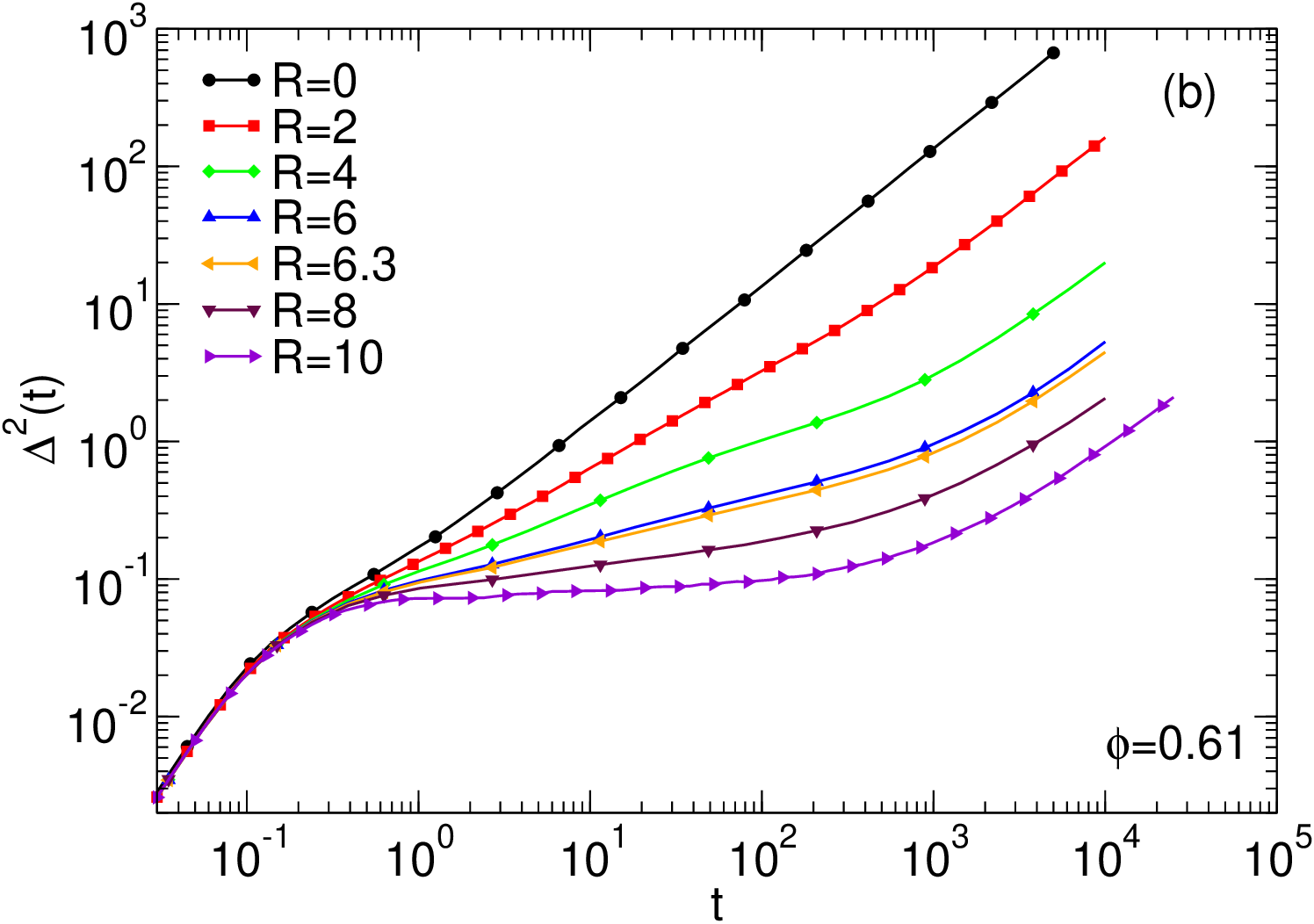}
\includegraphics[width=80mm,clip=true]{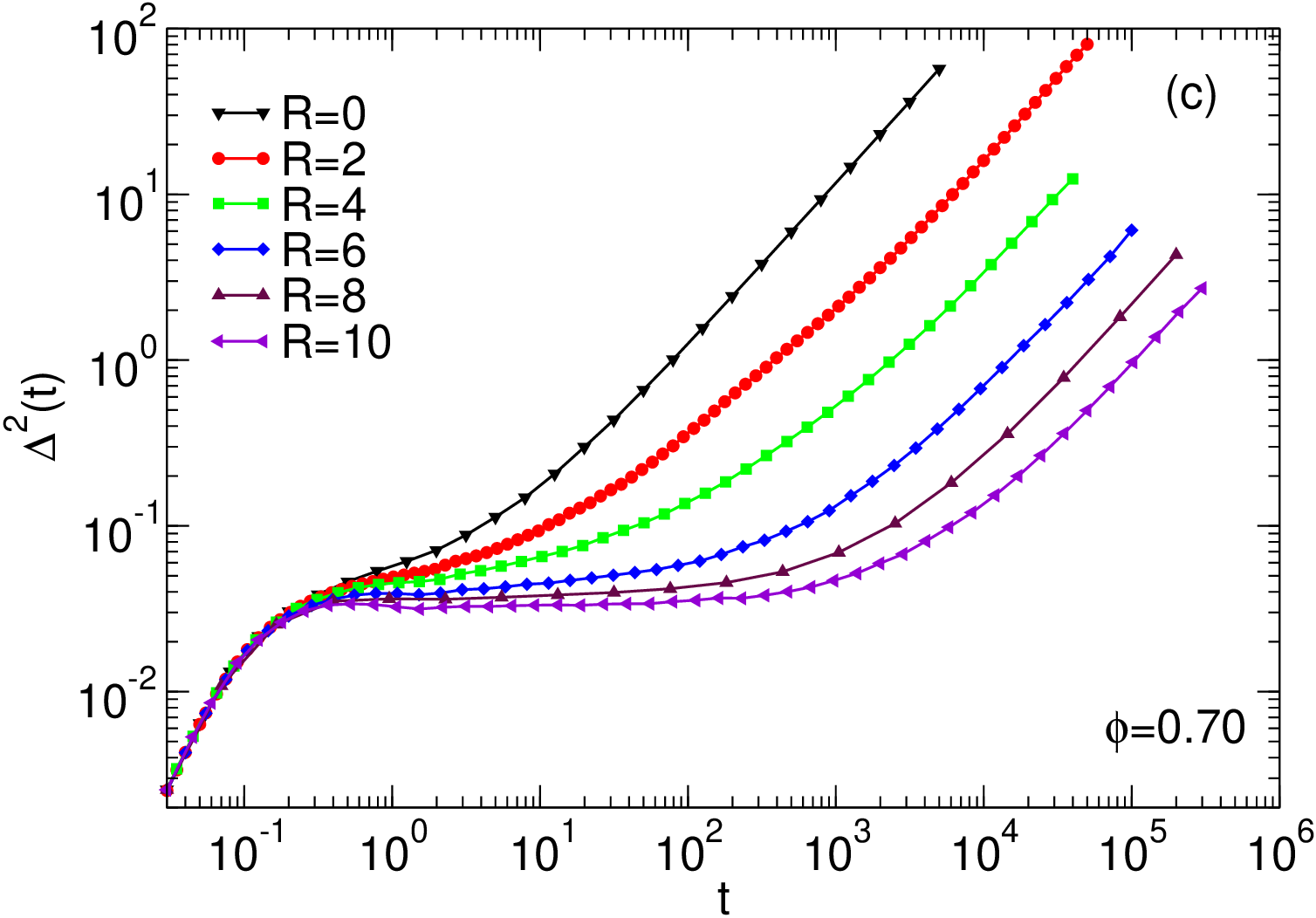}
\caption{(Color online) Variation of mean squared displacements
$\Delta^2(t)$ with  changing $R$ for (a)  $\phi= 0.50$, (b) $\phi=0.61$, and (c)
$\phi=0.70$, using a bond lifetime of $\tau_{\rm link}=10^2$. }
\label{msdt100}
\end{figure}

\begin{figure}[ht]
\includegraphics[width=80mm,clip=true]{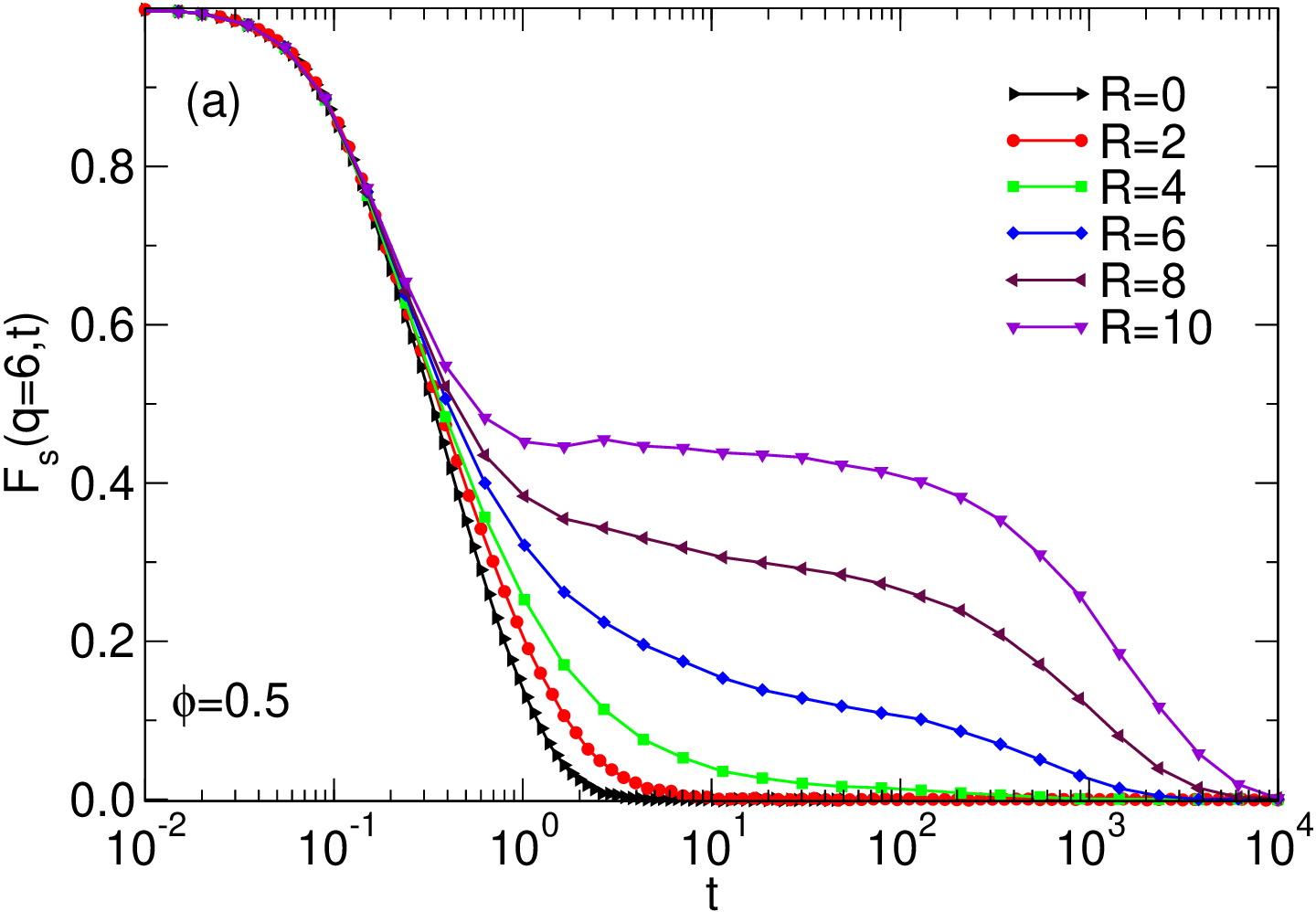}
\includegraphics[width=80mm,clip=true]{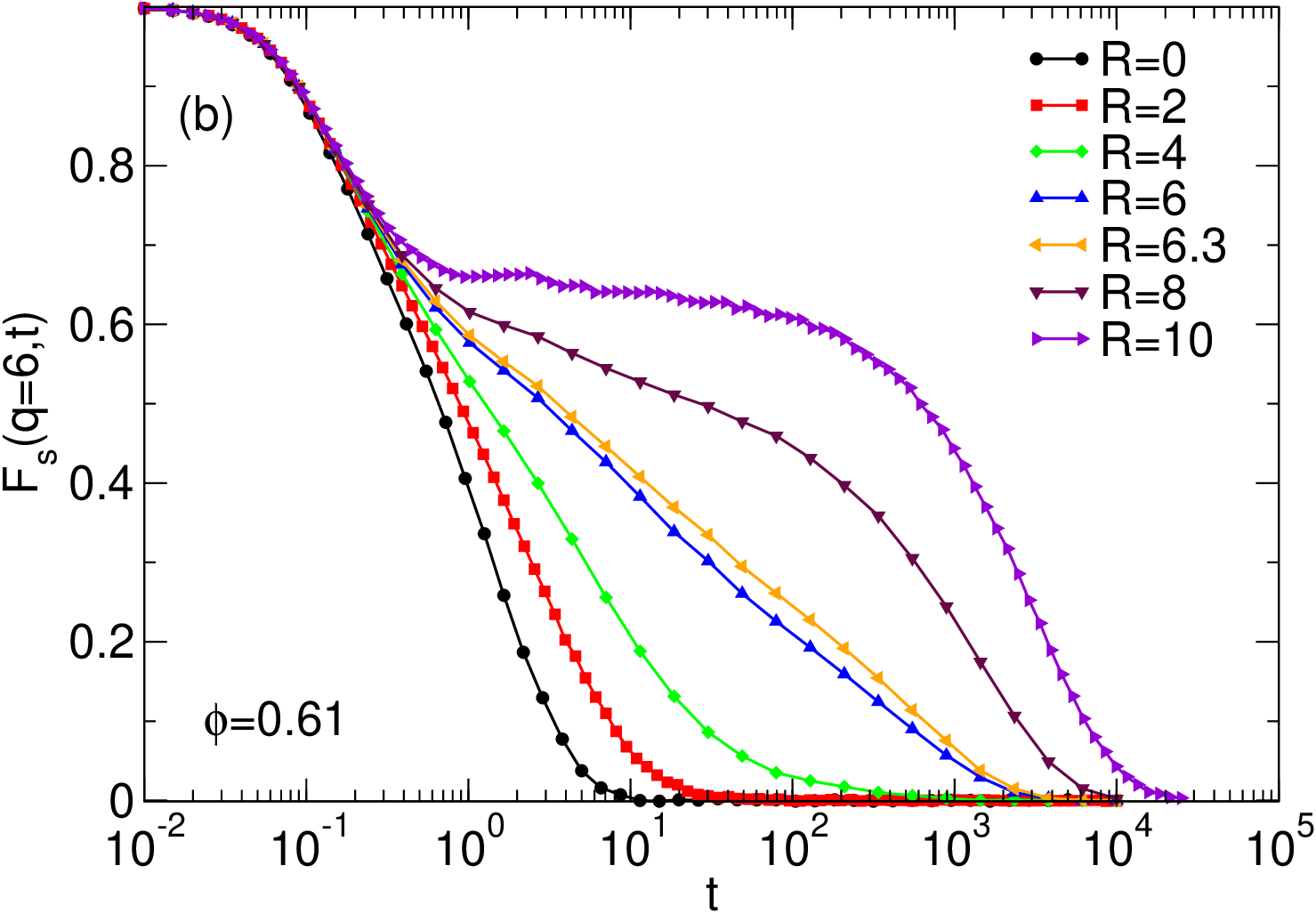}
\includegraphics[width=80mm,clip=true]{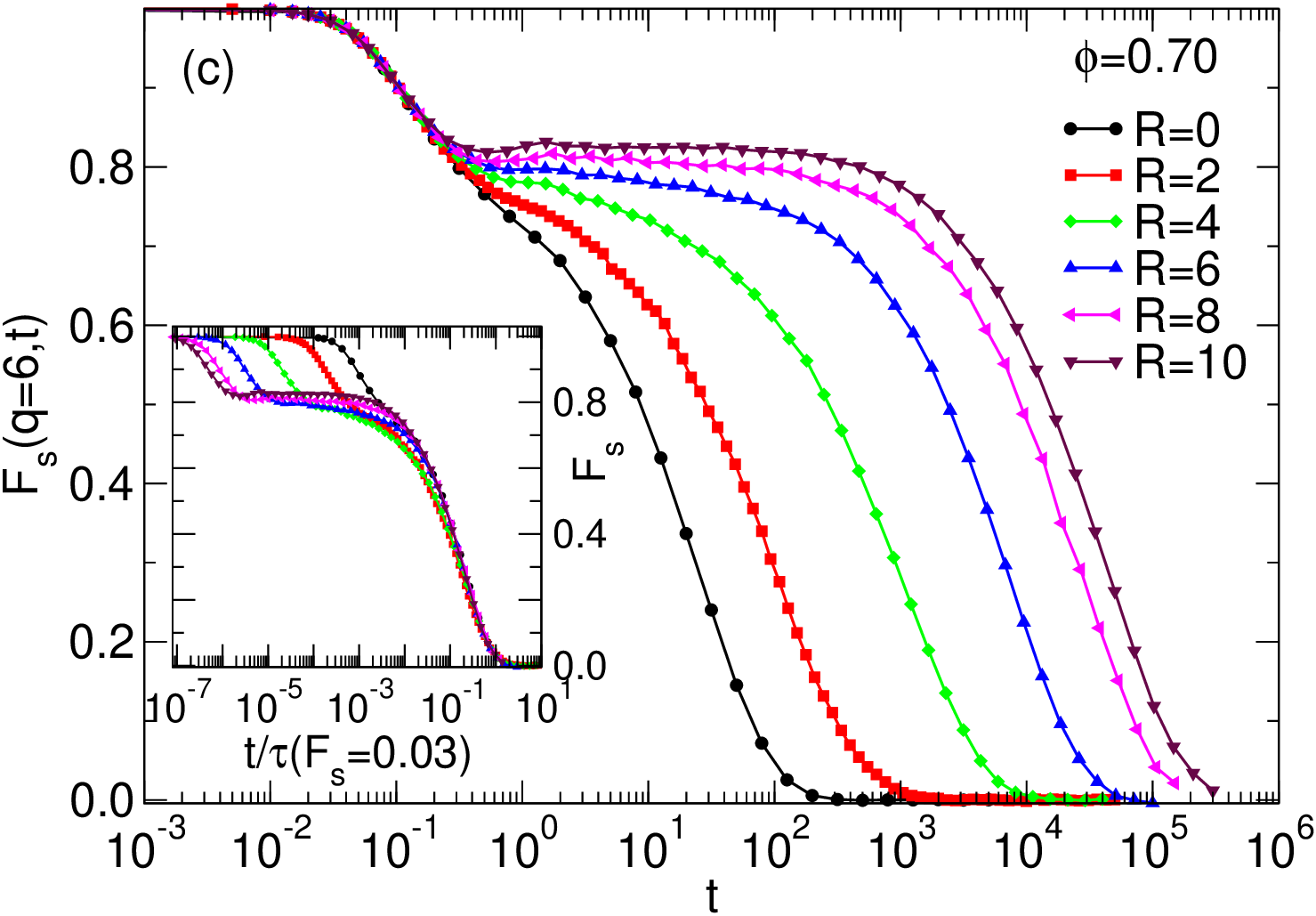}
\caption{(Color online) Variation of the self intermediate scattering
function with changing $R$.  (a)  $\phi= 0.50$, (b) $\phi=0.61$, and (c)
$\phi=0.70$, using $\tau_{\rm link}=10^2$.  Note the change of time span
in different panels, from a maximum time of $10^4$ in (a) to $10^6$ in
(c). The inset in panel (c) shows the collapse of correlation functions
for $\phi=0.70$ after rescaling by the relaxation time, similar to
time-temperature superposition principle.
}
\label{fsq6t100}
\end{figure}

\begin{figure}[h]
\includegraphics[width=80mm,clip=true]{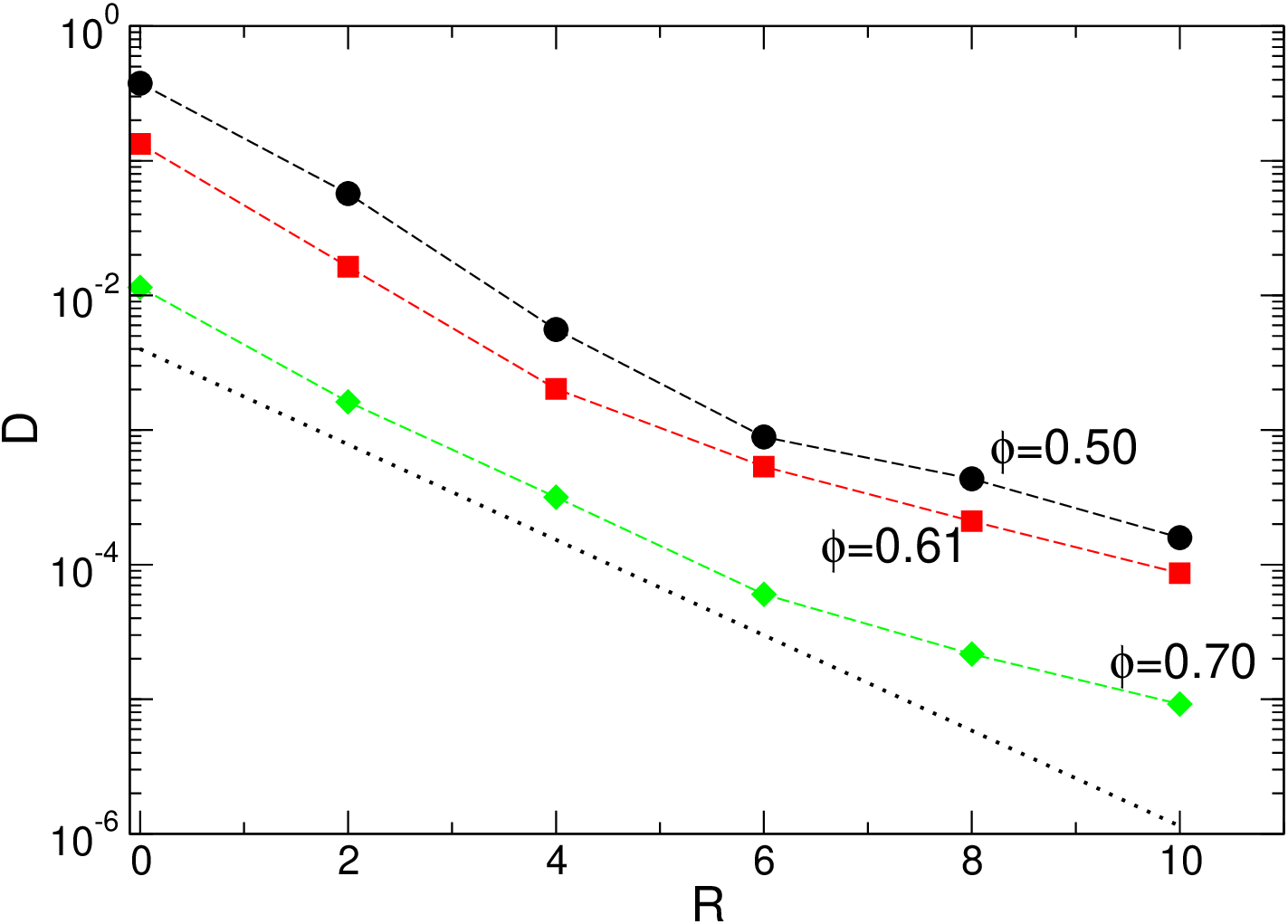}
\caption{(Color online) Variation of diffusion constant $D$
with R, for different $\phi$ (shown in Fig.~\ref{msdt100}). The dotted
line, corresponding to an exponential function, is drawn as a guide to the eye.}
\label{difft100}
\end{figure}

We now characterize the dynamical properties of the system by focusing on
two quantities: (i) the mean squared displacement, defined as $\Delta^2(t)
= \langle\frac{1}{N} \sum_i \left\langle | {\bf r}_i(t) - {\bf r}_i(0)
|^2 \right\rangle$ and (ii) the self-intermediate scattering function,
defined as $F_s(q,t) = \langle\frac{1}{N}\sum_j \exp \left( i {\bf q}
\cdot [ {\bf r}_j(t) -  {\bf r}_j(0) ] \right)\rangle$.  Here ${\bf
r}_i(t)$ is the position of particle $i$ at time $t$, $q$ is the
wave-vector, and $\langle .\rangle$ corresponds to the ensemble average.

For increasing volume fraction $\phi=0.50, 0.61, 0.70$, we discuss
simultaneously the data for $\Delta^2(t)$, shown in Fig.~\ref{msdt100},
and $F_s(q,t)$, shown in Fig.~\ref{fsq6t100}, computed at a wave-vector
value $q=6$. Thus, the measured $F_s(q,t)$ probes the relaxation
dynamics on length scales that are slightly larger than the average
particle diameter (the peak in the structure factor $S(q)$ occurs at
$q\approx{7.3}$).

We start in the pure gel phase ($\phi=0.50$). The mean squared
displacement, Fig.~\ref{msdt100}a, shows that the increasing number of
bonds restricts the motion of the particles in that for $R>0$ we see the
emergence of an intermediate regime which develops into a well-defined
plateau at $R=10$. The height of this plateau depends significantly on
$R$, showing that this cage motion is directly related to the transient
bonds between the particles. The self intermediate scattering function,
Fig.~\ref{fsq6t100}a, shows for small $R$ a very rapid decay. This
changes in that for $R$ around 4-6 a plateau develops at intermediate
and long times, the height of which depends strongly on $R$. The presence
of this increasing plateau height, which is reminiscent to the so called
type-A transition of mode-coupling theory~\cite{dawson}, indicates
that the relaxation mechanism is changing: For small $R$ the motion of
the particles is only weakly slowed down by the presence of the bonds,
which typically break on the time scale of $\tau_{\rm link}$. However,
for larger $R$ breaking a few bonds is not enough to allow the particles
to move since the remaining bond still allow to maintain the particle
inside its cage. Hence this makes that at large $R$ the relaxation
dynamics does not depend very strongly on $R$ anymore. This effect is
seen in Fig.~\ref{difft100} where we show the diffusion constant of the
particles, $D$, (as obtained from the mean squared displacement at long
times) as a function of $R$. For small $R$, $D$ shows a rather strong
$R-$dependence, whereas for $R>5$ this dependence becomes weaker. Below
we will discuss the $R-$dependence of the relaxation time in more detail.

Next, we look at the data for an intermediate density, viz. $\phi=0.61$,
and in Fig.~\ref{msdt100}b we show the corresponding $\Delta^2(t)$. Like
for $\phi=0.50$, the longtime diffusion decreases with increasing $R$,
and the $R-$dependence of the diffusion constant shows again a break at
around $R\approx 5$ (see Fig.~\ref{difft100}).  For this value of $\phi$
we observe, however, for $R>2$, at intermediate times a shoulder in
$\Delta^2(t)$. This shoulder, clearly visible for $R=4$, is related to
the presence of the bonds that lead to a caging of the particles on the
length scale related to $\ell$, the maximum extension of a bond. Thus
the hint of the short-time plateau (when $\Delta^2(t)\approx{0.1}$)
and again one at later time (when $\Delta^2(t) \approx 1$) reflects
the presence of the two different mechanisms for constraining
particle motion, viz. local steric hindrance and the network bonds.
Since each type of caging leads to a plateau in the intermediate
scattering function~\cite{pre2010}, the existence of the two competing
mechanisms makes that at intermediate times $F_s(q,t)$, shown in
Fig.~\ref{fsq6t100}b, has a very slow, almost logarithmic, decay,
if $R$ is around 6. More details on this particular case are give in
the context of Fig.~\ref{decon}. If one compares the data for $R=10$
at the two volume fractions $\phi=0.50$ and $\phi=0.61$, we recognize
that the height of the plateaus in $F_s(q,t)$ increases with $\phi$
from which one can conclude that the proximity of the particles leads
to increased tightening of the cage.  

If the density is increased further to $\phi=0.70$, the relaxation
dynamics becomes strongly dominated by the steric hindrance
mechanism. Already for $R=0$ one sees a weak plateau in $\Delta^2(t)$
and its length grows rapidly with increasing $R$ without changing much
its height (see Fig.\ref{msdt100}c). This is the typical behavior of
simple glass-forming liquids~\cite{kob_95}. At the same time the self
intermediate scattering function shows the growth of a shoulder with
{\it finite} height and this height depends again only weakly on $R$,
Fig.~\ref{fsq6t100}c. In contrast to the case at lower densities,
here the shape of the correlator is basically independent of $R$. This
is demonstrated in the inset of Fig.~\ref{fsq6t100}c where we plot
$F_s(q,t)$ as a function of $t/\tau$, with the relaxation time $\tau$
defined by $F_s(q,\tau)=0.03$. The fact that this presentation of the
curves leads to a nice master curve shows that we have for this system a
time-$R$ superposition, in analogy to the time-temperature superposition
found in simple glass-forming systems~\cite{book}.

Despite the qualitative changes seen in $\Delta^2(t)$ and $F_s(q,t)$
if $\phi$ is increased, the $R-$dependence of the diffusion constant
for $\phi=0.70$ is very similar to the one seen at lower densities,
see Fig.~\ref{difft100}. Also for this high value of $\phi$ we see that
this dependence is relatively strong at small $R$ and becomes weaker if
$R>6$. Hence the fact that at low $R$ just few bonds have to be broken
in order to allow a particle to move whereas at high $R$ this is not a
sufficient condition, is reflected in the $R-$dependence of $D$ at {\it
all} $\phi$.


\subsection{Varying the bond life-time $\tau_{\rm link}$}

\begin{figure}[]
\includegraphics[width=80mm,clip=true]{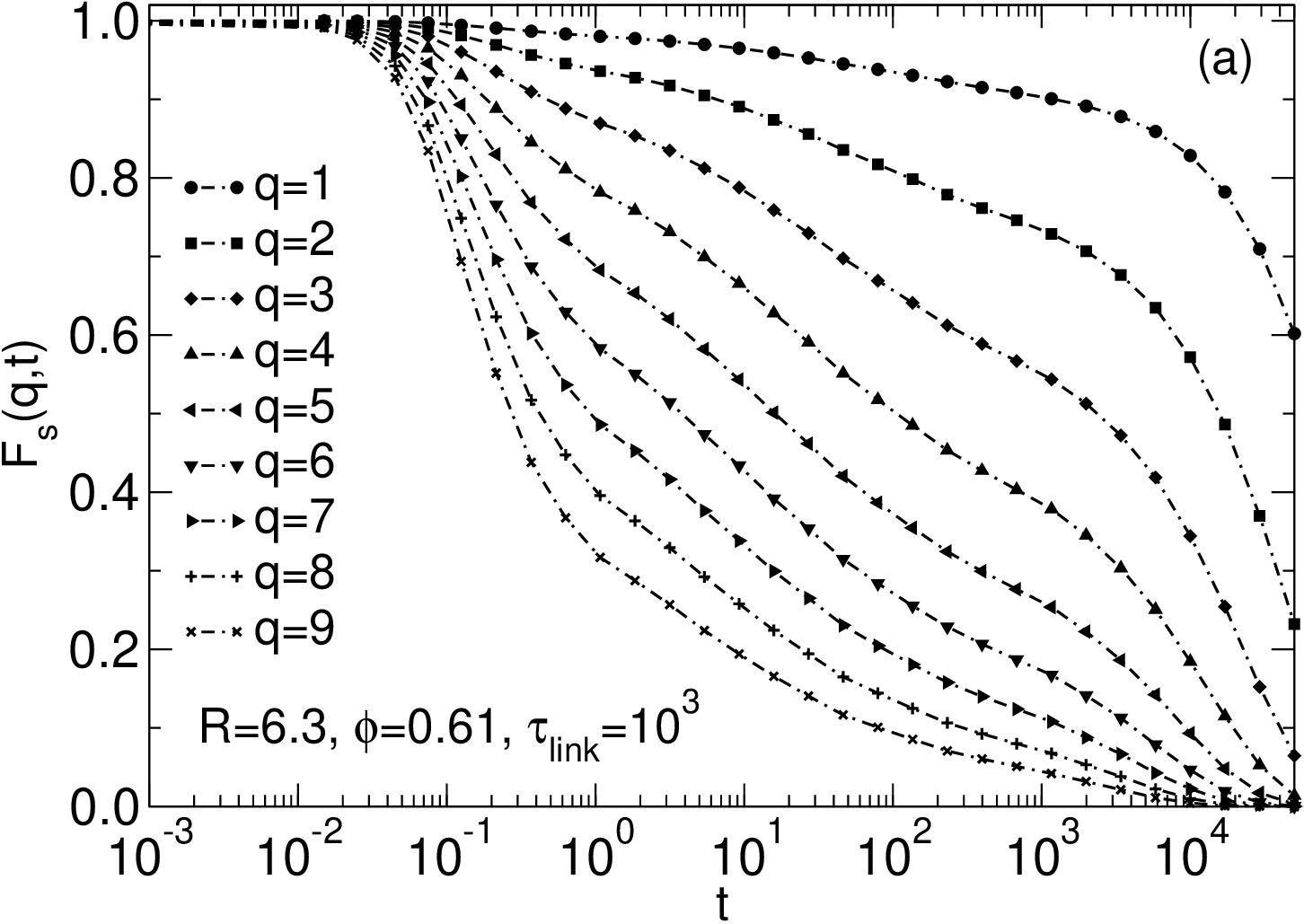}
\includegraphics[width=80mm,clip=true]{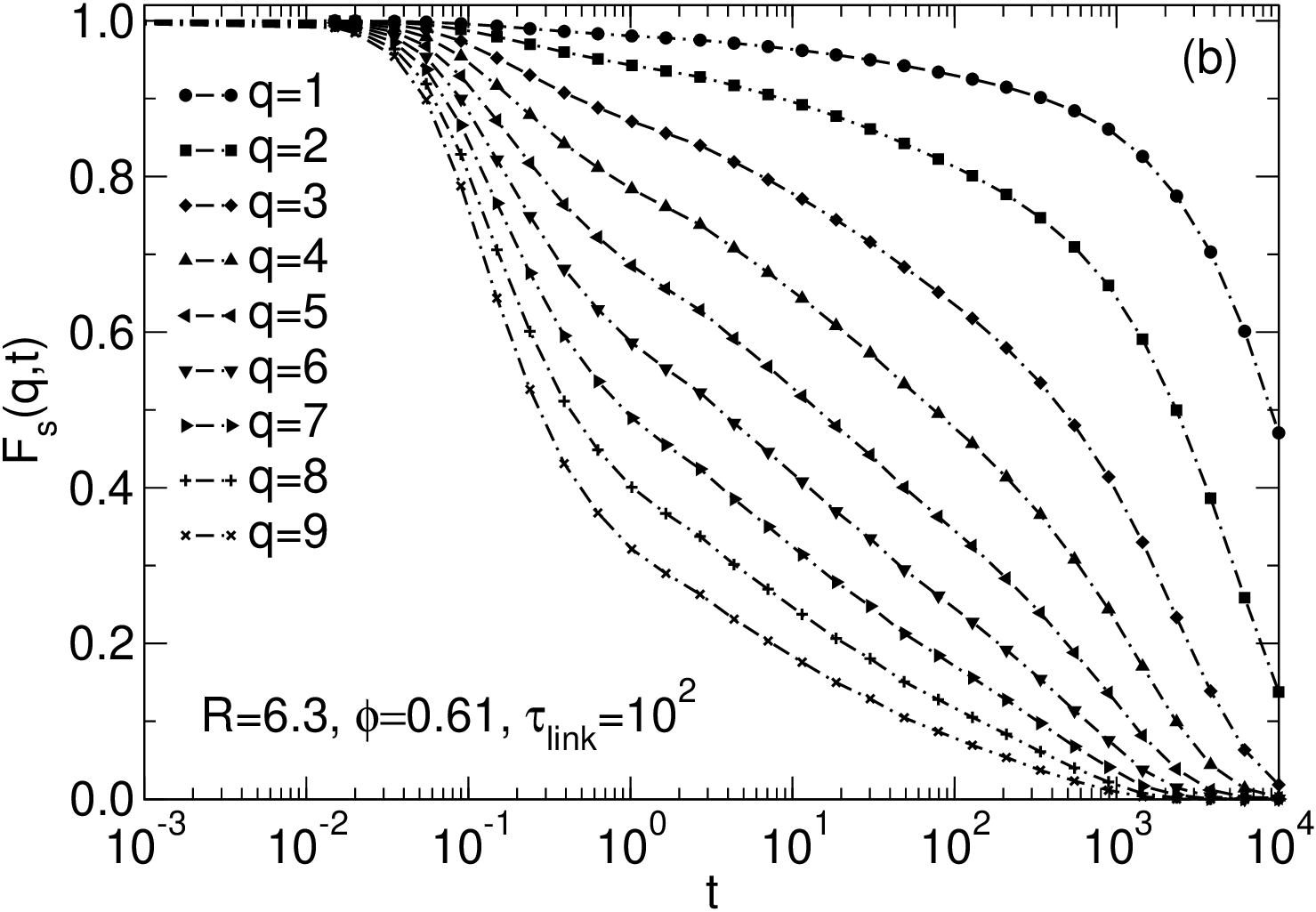}
\includegraphics[width=80mm,clip=true]{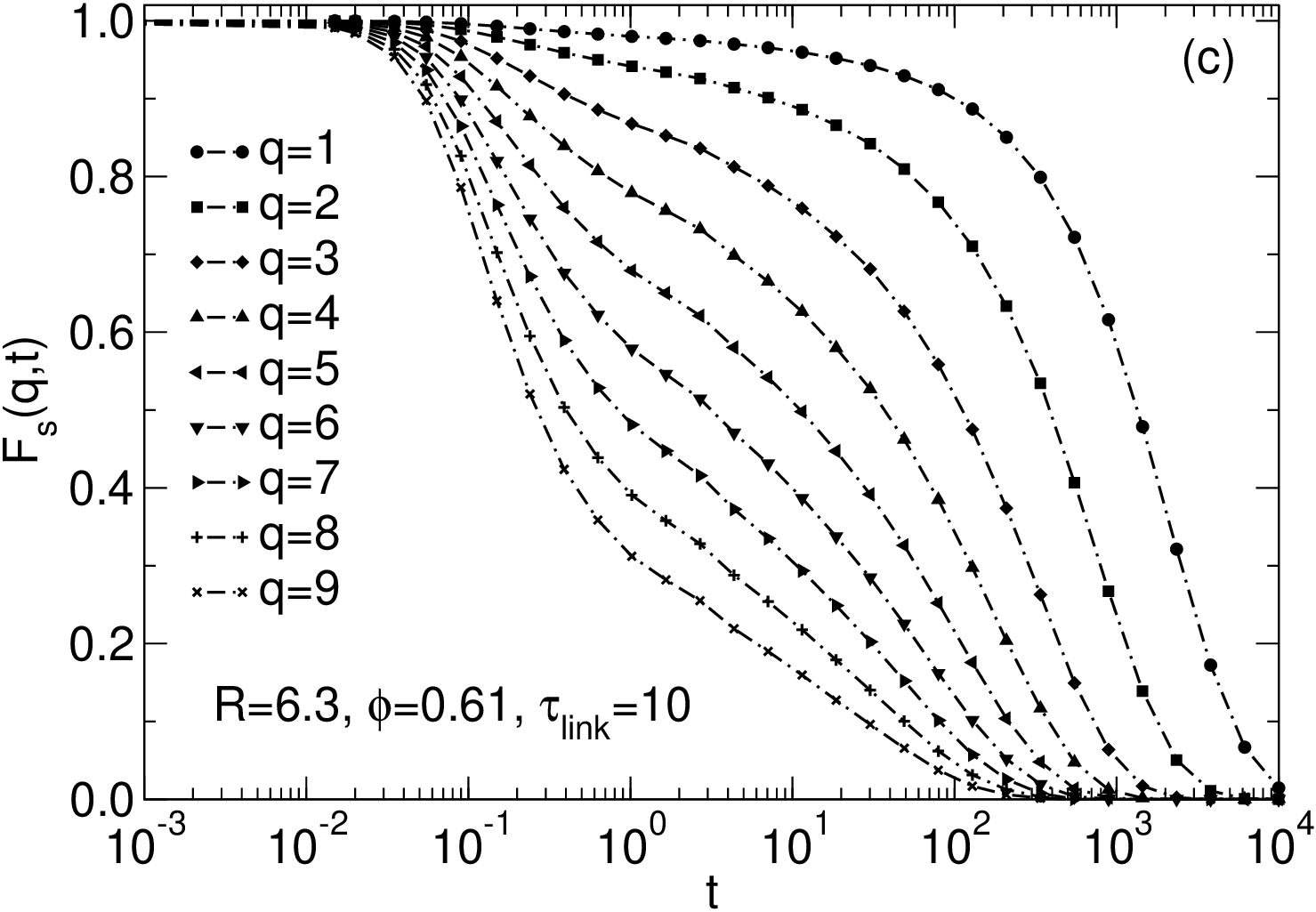}
\caption{(Color online) $F_s(q,t)$ for $R=6.3, \phi=0.61$, a) $\tau_{\rm link}=10^3$ b) $\tau_{\rm link}=10^2$
and c) $\tau_{\rm link}=10$.}
\label{r63pv61fsqall}
\end{figure}

We will now explore further the interplay between the two processes
leading to slow relaxation, i.e. the nearest-neighbor caging and
the constrained motions due to the polymer bonds. Since the relative
importance of these two processes depends on the lifetime of the polymer
bonds, we will in the following vary this lifetime and consider values
of $\tau_{\rm link}=10, 10^2$, and $10^3$, and at fixed state-point of
$R=6.3,\phi=0.61$, where logarithmic decay in the time-correlation
function is observed. We study how the shape of $F_s(q,t)$ changes with
varying $\tau_{\rm link}$ and will relate this to the interplay between the
two processes.  This is done for different values of wave-vector in order
to see the how relaxation timescales vary over different lengthscales.

We begin by looking at the case of the large $\tau_{\rm link}=10^3$,
i.e. when the polymer bonds hinder the motion on a time scale longer
than the steric hindrance effect, see Fig.~\ref{r63pv61fsqall}a. The
correlation function reflects three different relaxation processes
which can be seen for all value of $q$. Initially, the particles rattle
inside the cage, resulting in partial relaxation of the correlation
function on a timescale $\tau_\beta \approx 1$. Later on, the particles escapes
from the cage of neighboring particles, (which for this value of $\phi$
not very pronounced), but the relaxation process is then held up by the
polymer bonds. 
Eventually, the polymer network rewires on a timescale which is proportional to 
$\tau_{\rm link}=10^3$ and the particles start the final relaxation process.
The height of the plateau in $F_s(q,t)$ increases with decreasing $q$,
which is the typical behavior for a glassy system \cite{kob_95,nauroth}.
However, for the range of wave-vectors explored, the final  timescale for
decay of $F_s(q,t)$ depends only weakly on $q$. Thus, over these length
scales, the relaxation process is determined by the reconfiguration
of the network. However, at larger length scales, one can expect that
hydrodynamic effects will eventually dominate and this will then determine
the relaxation timescales.

For intermediate values of the bond lifetime, $\tau_{\rm link}=10^2$
in Fig.~\ref{r63pv61fsqall}b, the final decay has moved to shorter
times and makes that now the interplay between the two processes
results in a logarithmic decay of $F_s(q,t)$ as discussed elsewhere
\cite{pre2010}. This logarithmic dependence is seen for a range of $q$
values (see Fig.~\ref{r63pv61fsqall}b), with the time-window over which
it exists decreasing with decreasing $q$.  This implies that this form
for the correlation function occurs only for specific combination of
the relaxation timescales of the two processes of steric hindrance and
eventual network relaxation. At sufficiently small $q$, hydrodynamics
makes that the relaxation becomes so slow that the second relaxation step
is no longer visible out and hence the logarithmic $t-$dependence is no
longer observed. Finally we mention that the logarithmic shape in the
relaxation function as discussed here is not related to any underlying
higher order mode-coupling transition, in contrast to the case of
certain colloidal systems for which similar relaxation functions have
been observed \cite{zaccarelli3}.

If we set $\tau_{\rm link}$ to a small value, this three-step relaxation
can no longer be observed, as is seen for the case of $\tau_{\rm link}=10$
in the bottom panel of Fig.~\ref{r63pv61fsqall}: For all $q$'s the
curve show a (seemingly) simple two step relaxation, since the third step
(related to the bonds) starts already when the second step (related to
steric hindrance) is not yet completed and hence the two processes become
completely mixed in time. Below we will briefly come back to this effect.

\subsection{When the bonds are permanent}

\begin{figure}[ht]
\includegraphics[width=80mm,clip=true]{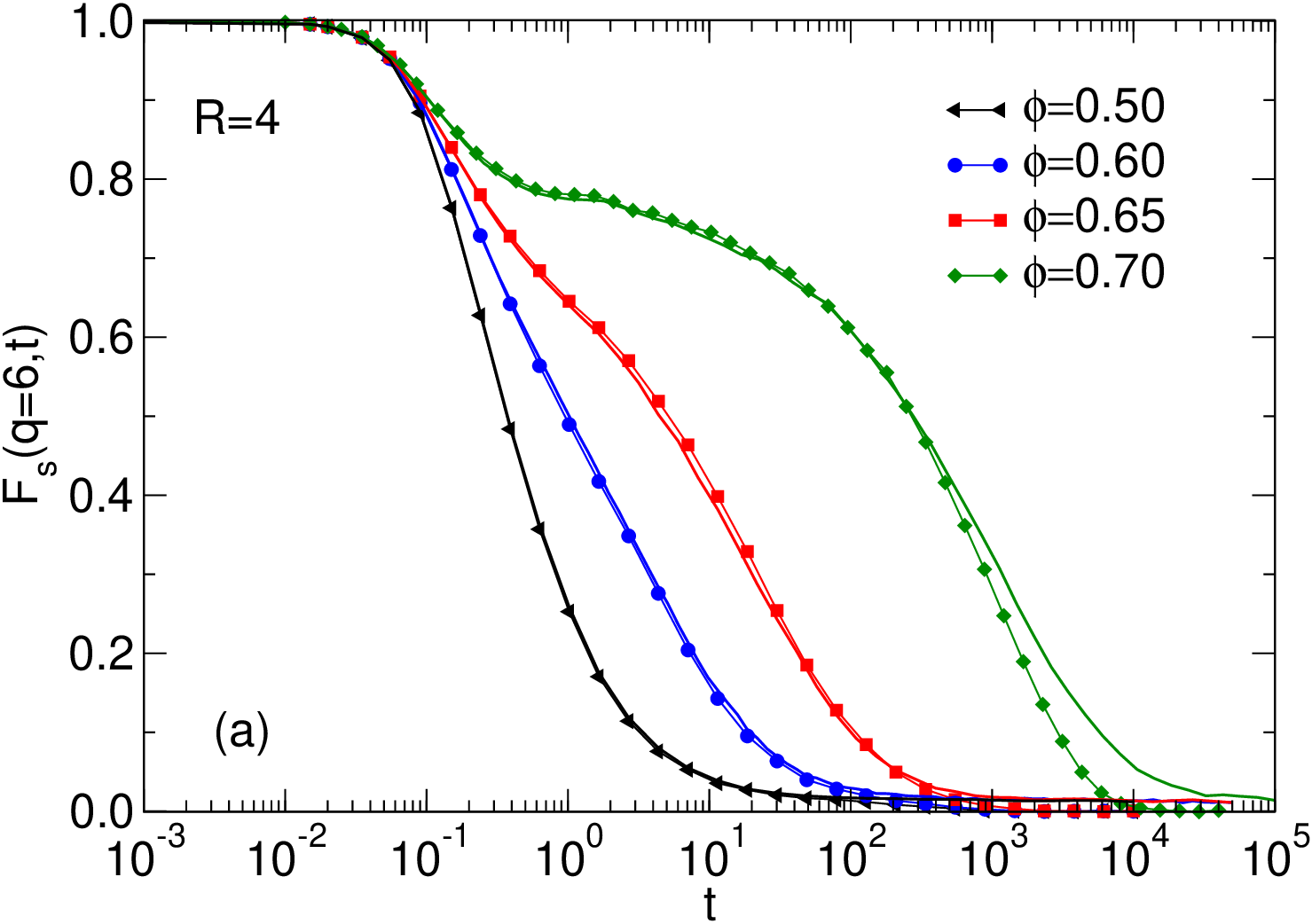}
\includegraphics[width=80mm,clip=true]{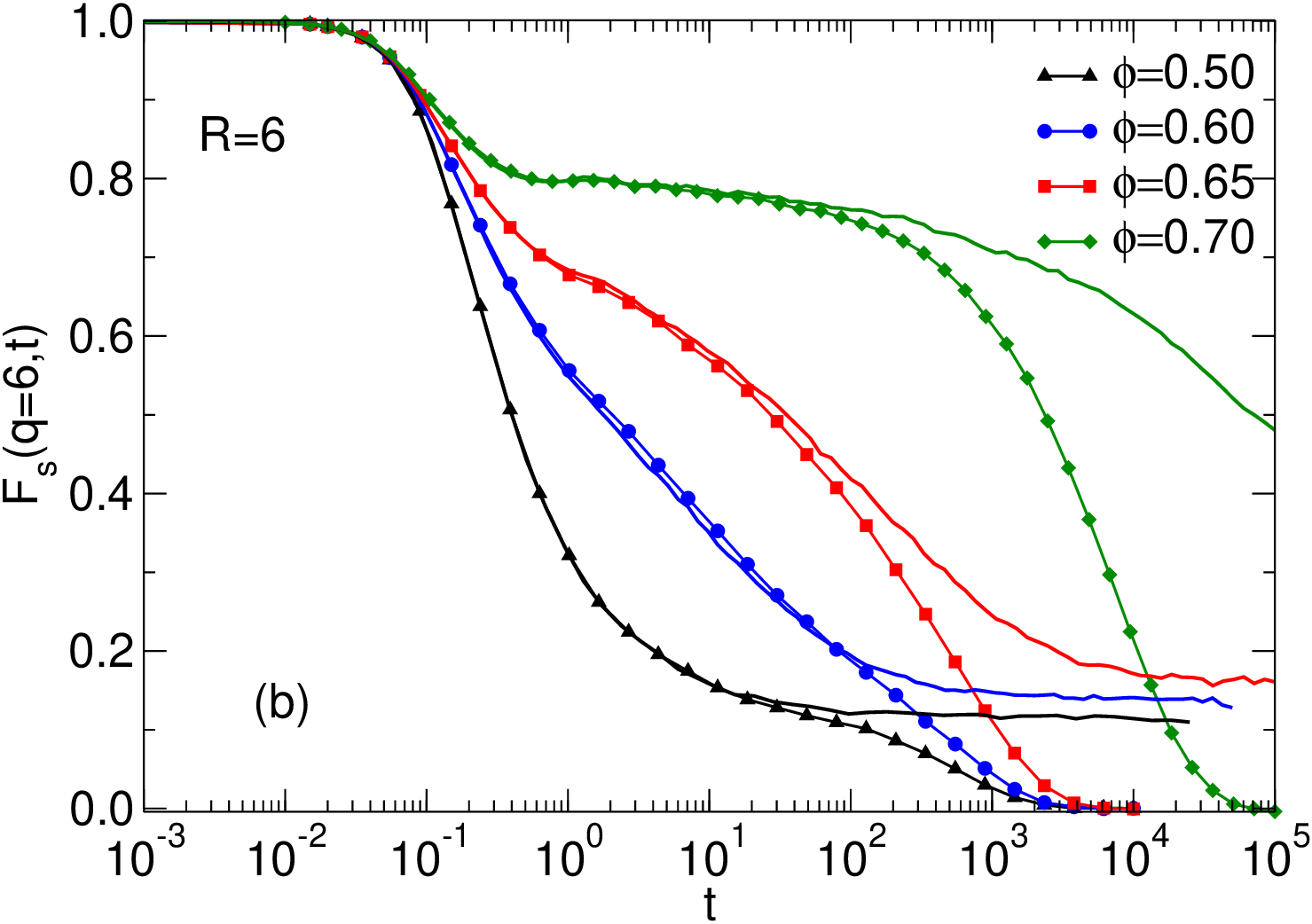}
\includegraphics[width=80mm,clip=true]{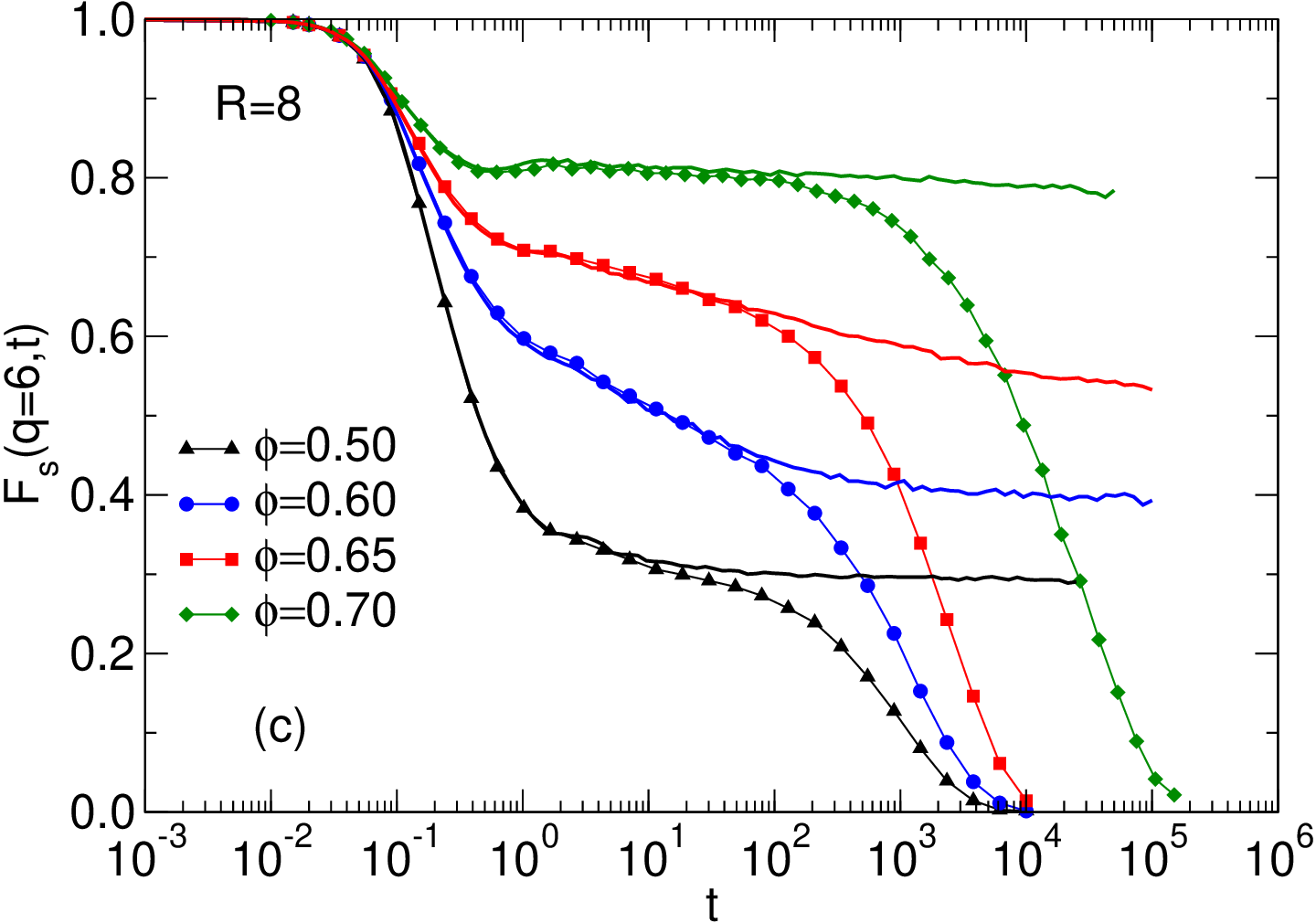}
\caption{(Color online) 
$F_s(q,t)$ for different $\phi$ and $R=4,6,$ and 8 when the 
bond lifetime is $\tau_{\rm link}=100$ (lines
with symbols) and when $\tau_{\rm link}=\infty$ (lines). The wave-vector is $q=6.0.$}
\label{fsq6t100tinf}
\end{figure}

A useful way to check the influence of the polymer bonds on the relaxation
dynamics is by comparing the relaxation functions for the case when the
bonds are permanent, i.e.~$\tau_{\rm link}=\infty$, to those when the
lifetime  is finite. In Fig.~\ref{fsq6t100tinf}, we do this comparison for
different connectivities ($R=4,6,$ and 8) and different values of $\phi$.

In Fig.~\ref{fsq6t100tinf}a, we show $F_s(q=6,t)$ for the case of
$R=4$ with the two different bond lifetimes $\tau_{\rm link}=10^2$
and $\tau_{\rm link}=\infty$.  For $\phi=0.50,$ 0.60, and 0.65,
the timescales for overcoming the steric hindrance are the same for
both lifetimes. However, for all $\phi$, we find that the correlation
function for $\tau_{\rm link}=\infty$ shows a plateau at long times (not
visible in this plot), which is due to the fact that the frozen bonds
in the percolating gel-network prevent the complete relaxation of the
system. In contrast the curves for the finite $\tau_{\rm link}$
vanish at long times.  For small and intermediate $\phi$ the two
sets of curves are very similar, indicating that the presence of a few
bonds does not change the dynamics significantly. Only for $\phi=0.7$ one
sees a substantial difference in that the correlator for the permanent
links decays slower than the one with $\tau_{\rm link}=10^2$. It is
reasonable that these differences are noticeable for times somewhat longer
than $10^2$, i.e. the time scale of $\tau_{\rm link}$.

If we increase the connectivity to $R=6$ (Fig.~\ref{fsq6t100tinf}b),
we see that the behavior is qualitatively similar to $R=4$ in that for
all values of $\phi$ the two curves track each other up to times around
$10^2$, i.e. the time of the finite $\tau_{\rm link}$. For larger
times the correlators for $\tau_{\rm link}=10^2$ decay to zero whereas the
ones for $\tau_{\rm link}=\infty$ show at long times a marked plateau.
The height of this plateau depends now more strongly on $\phi$ than it was
the case for $R=4$, showing that if $R$ is increased the life time of the
bonds becomes more influential.  This is reasonable since it is related
to the general observation that in glass-forming system small changes influence
the relaxation dynamics increasingly more the slower the dynamics is.
We also note that for $\phi=0.70$ the correlator for the permanent bonds
becomes very stretched.  This sluggish relaxation might be related
to the fact that for this value of $R$ there are, in addition to the
percolating cluster, clusters of different sizes (see Ref.~\cite{pablo2}
for typical distributions), thus giving rise to relaxation dynamics that
spans many orders of magnitude in time and hence to a very stretched
average correlation function. The stretching of the correlator for
the frozen bonds could, however, also be due to the fact that these
different clusters hinder each other resulting in the overall slowdown
of the dynamics \cite{puertas2}.

Next, we increase the number of bonds even more, viz. $R=8$, as shown
in Fig.~\ref{fsq6t100tinf}c. We see that for the case of permanent
bonds the height of the asymptotic plateau has increased strongly
in comparison to the case of $R=6$. As a result the correlator for
$\phi=0.70$ shows only a negligible decay of the correlation function
if the bonds are permanent. The motion is so constrained by these bonds
that the height of the asymptotic plateau, caused by the permanent
bonds, becomes comparable to the one related to the steric hindrance.
Also for $\phi=0.5$ the height of the second plateau has increased so
much that the relaxation from caging is now completely masked. However,
for the intermediate values of $\phi$, one does notice a difference
between the two plateau heights and the correlations functions decay in
a very stretched fashion from one to the other. In fact, the decay is
so slow that the time-dependence is seen to be logarithmic (nearly for
five decades in the case of $\phi=0.65$).  Note that this logarithmic
decay is due to the heterogeneous relaxation of the floppy clusters of
frozen bonds, which we will discuss later in further detail. Thus this
mechanism is different from the one leading to the logarithmic relaxation
seen in Fig.~\ref{fsq6t100} which was due to a unique combination of
the two {\it finite} relaxation timescales.

\begin{figure}[]
\includegraphics[width=70mm,clip=true]{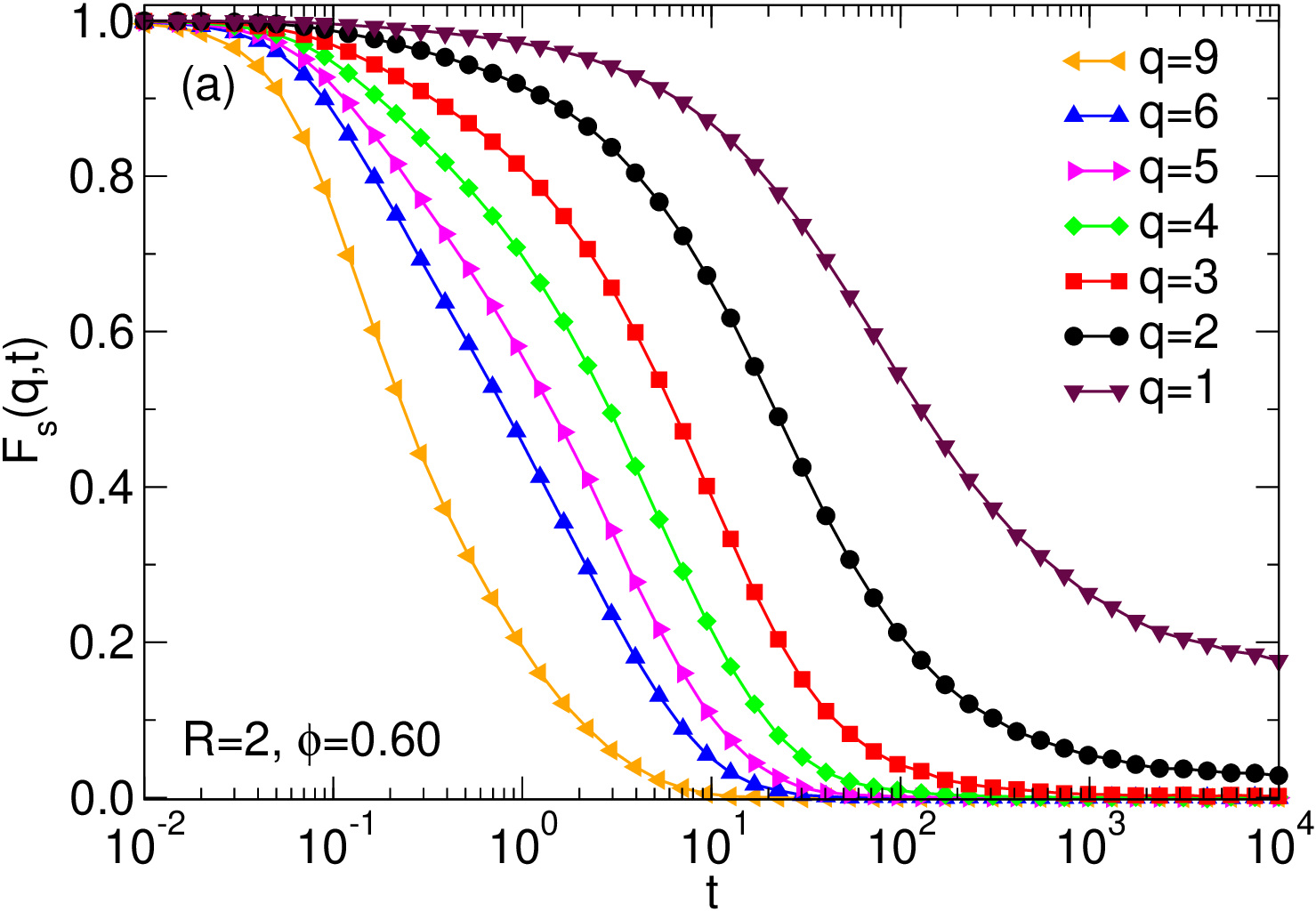}
\includegraphics[width=70mm,clip=true]{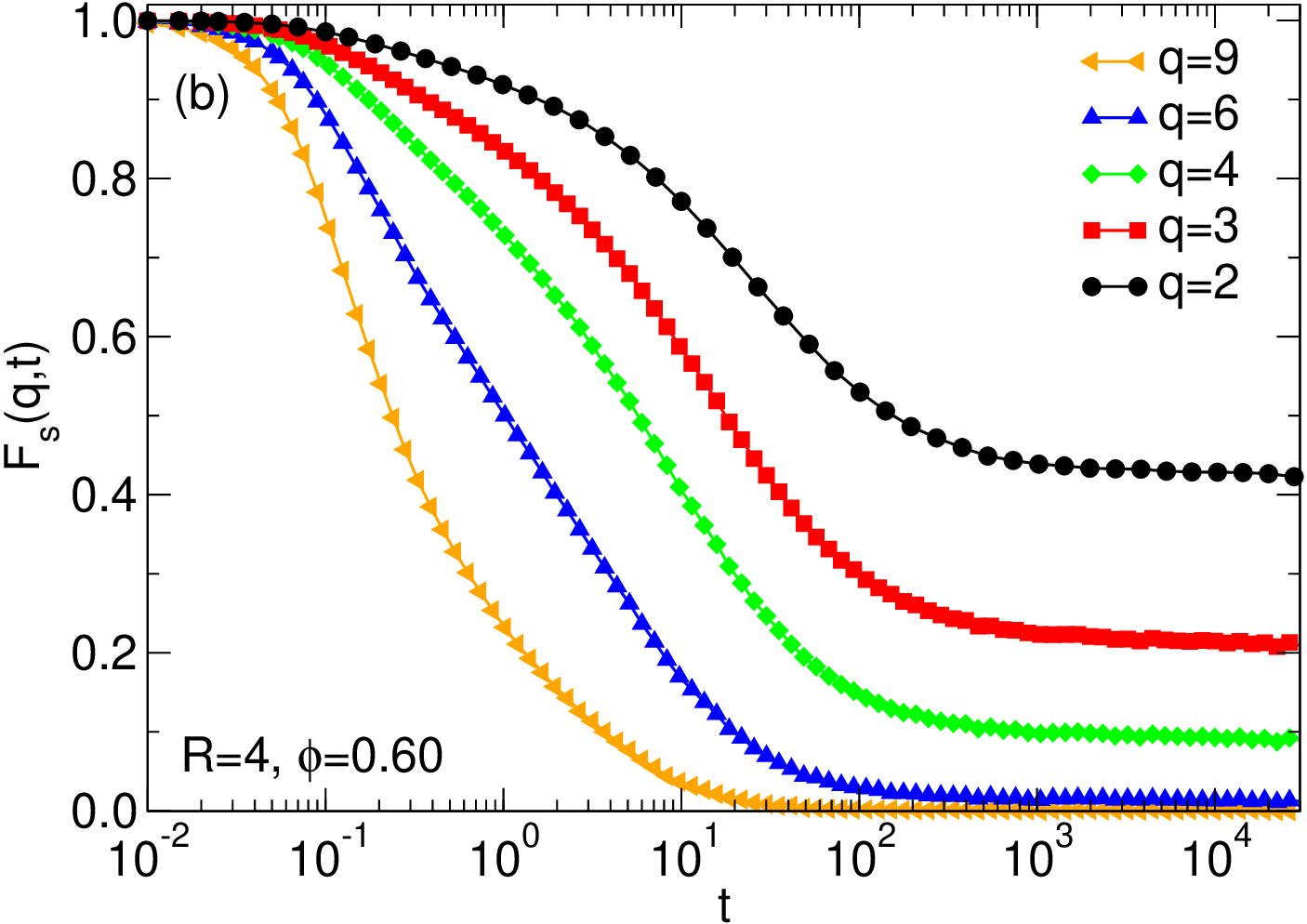}
\includegraphics[width=8cm,clip=true]{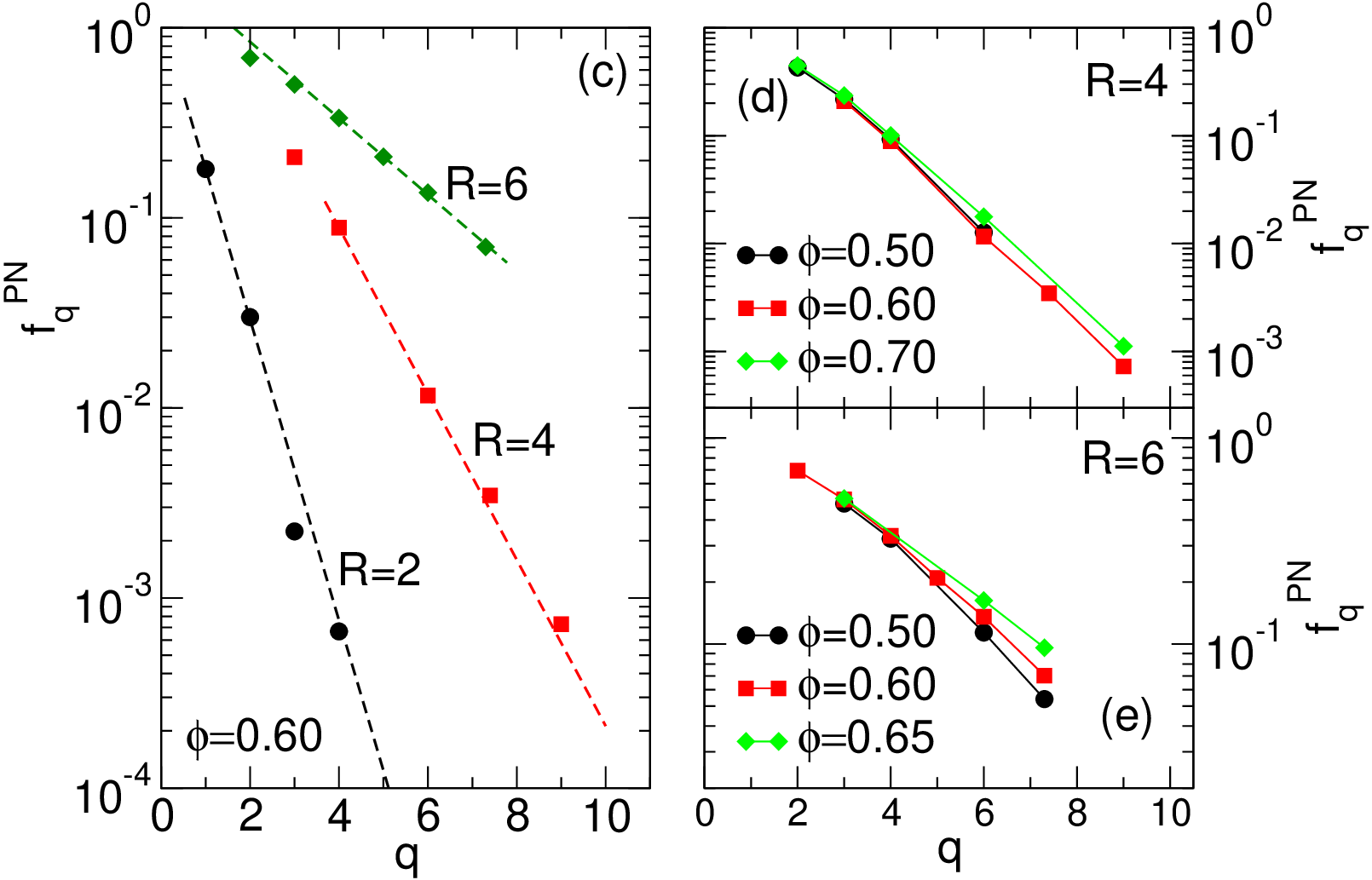}
\caption{(Color online) (a),(b): Variation of $F_s(q,t)$ with wave-vector
$q$ for $R=2$ and 4 at $\phi=0.60$ with $\tau_{\rm link}=\infty$.  (c):
Height of plateau, $f_q^{\rm PN}$, of $F_s(q,t)$ at long times as a function
of $q$ for $R=2$, 4, and 6.  The dashed lines are fits of the form
$\exp{(-q/\xi)}$ with the corresponding $\xi=0.55$, 0.99, and 2.27.
(d),(e): $f_q^{\rm PN}$ vs. $q$ for different values of $\phi$ for $R=4$ and 6. }
\label{ap1}
\end{figure}

We now study the floppiness of this network of particles connected by the
permanent bonds by probing the wave-vector dependence of the relaxation
functions $F_s(q,t)$. In Figs.~\ref{ap1}a and b, we show for $\phi=0.60$
the variation of $F_s(q,t)$ for $R=2$ and 4, i.e. in the region of
the phase diagram where gelation sets in. The height of the plateau at
long times, also called non-ergodicity parameter, is a measure for the
stiffness of the network on the length scale $q$ considered. Comparing
the two panels we recognize that, for a given $q$, the height of the
plateau increases with increasing $R$. Denoting this height by $f_q^{\rm
PN}$, we show in Fig.~\ref{ap1}c that $f_q^{\rm PN}$ shows basically
an exponential decrease in $q$ with a slope that decreases rapidly
with increasing $R$. That $f_q^{\rm PN}$ decreases with increasing $q$
is of course reasonable since on small length scales the particles have
more leeway to flop around than on large length scales. Note, however,
that this exponential dependence is in contrast to the one found for
the height of the plateau due to the steric hindrance, the latter being
basically a gaussian function~\cite{nauroth}.  Since in the representation
of Fig.~\ref{ap1}c such a gaussian dependence is given by a parabola,
we see that such a curve will intersect the one for $f_q^{\rm PN}$ at
a certain value of $q_x$. For $q< q_x$ the plateau due to the steric
hindrance is above $f_q^{\rm PN}$, thus making that one observes two
plateaus in the correlator. However, for $q>q_x$ the plateau at long
times is higher than the steric one, thus making that the latter one
will be completely masked by the former and thus the correlator will
show only one plateau.

We have also studied how the $q-$dependence of $f_q^{\rm PN}$ changes with
the volume fraction and in Figs.~\ref{ap1}d and e, we show $f_q^{\rm PN}$
for $R=4$ and 6, respectively. For both cases we see that the rigidity
of the network at large scales, i.e.~small $q$, is not affected by the
volume fraction which is not surprising. For $R=4$ we see that this is
also true at small length scales whereas for $R=6$ we note a significant
$\phi-$dependence if $q$ is large. This difference is likely related
to the fact that the system with $R=6$ is much more sluggish than the
one for $R=4$, see Fig.\ref{fsq6t100tinf}, and hence small changes
(here in $\phi$) will have a stronger impact on the dynamics.

\begin{figure}[h]
\includegraphics[width=80mm,clip=true]{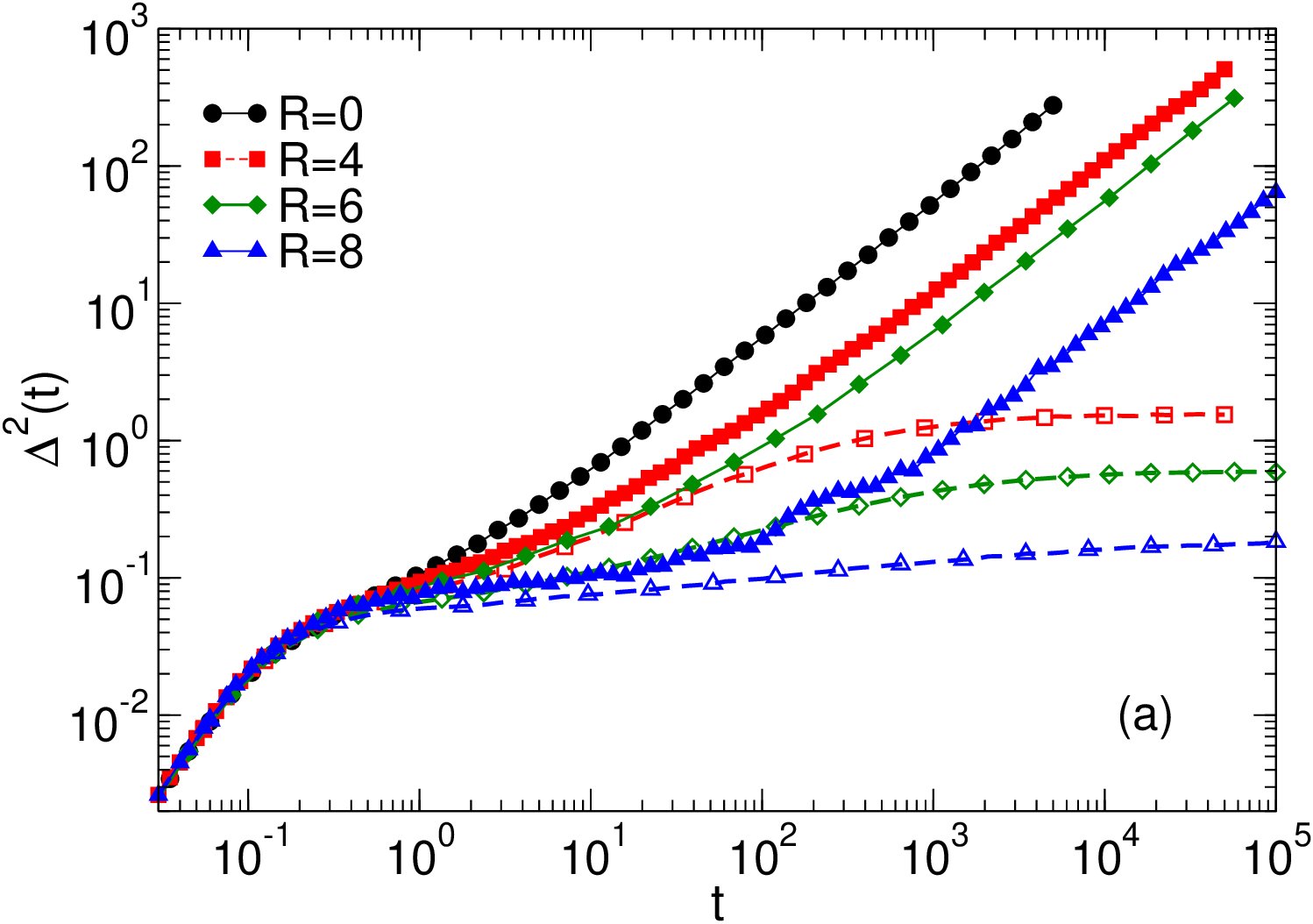}
\includegraphics[width=80mm,clip=true]{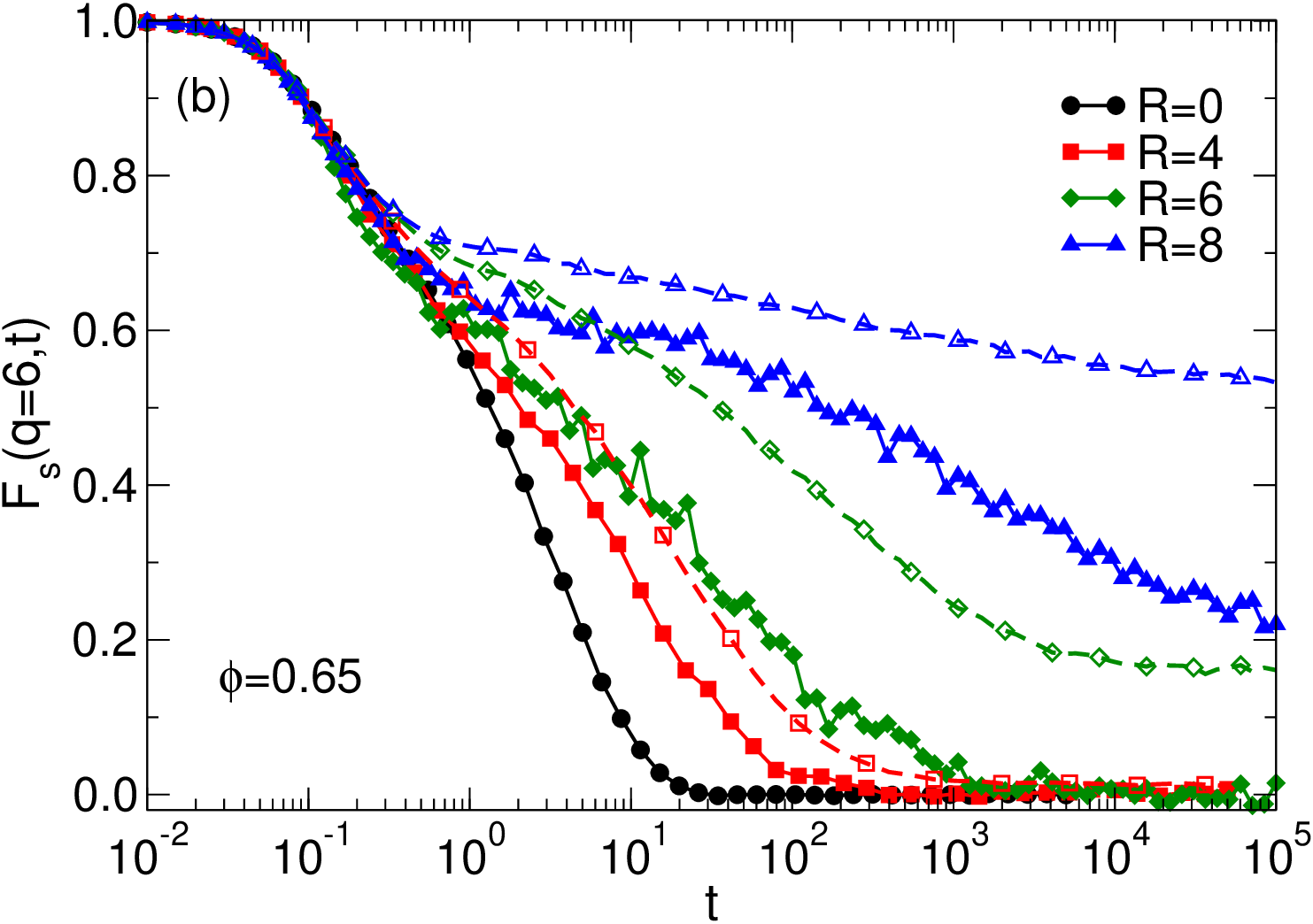}
\caption{(Color online) Relaxation dynamics of the clustered (dashed lines
with open symbols) and non-cluster particles (full lines with filled
symbols) for the case of frozen bonds (i.e $\tau_{\rm link}=\infty$),
at $R=4$, 6, and 8 for volume fraction $\phi=0.65$. (a): Mean squared
displacement $\Delta^2(t)$. (b) Self intermediate scattering function
$F_s(q,t)$ for $q=6$.}
\label{figclustfree}
\end{figure}

Finally we disentangle the dynamics of the particles that belong to the
percolating cluster from the one that are not attached to it. In the
following discussion these are referred as clustered and non-clustered
particles, respectively. The objective is to clarify the respective
contributions to the different dynamical quantities that we have
discussed above. We do this comparison for an increasing number of
connections $R$ at a fixed (large) volume fraction of $\phi=0.65$.
In Fig.~\ref{figclustfree}, we show the data for $\Delta^2(t)$ and $F_s(q,t)$ 
for these two families of particles.

The mean squared displacement of the clustered particles shows after the
ballistic regime at short times a shoulder that is related to the cage
of the steric hindrance (Fig.~\ref{figclustfree}a). This localization
is, however, only temporary and is followed by a further increase of
$\Delta^2(t)$. Only at longer times $\Delta^2(t)$ saturates at a height
that depends on $R$. We note that the approach to this asymptotic
height becomes increasingly slow with increasing $R$ and in fact for
$R=8$ the time dependence is close to logarithmic and does not end
within the time window of our simulation.  This behavior
is also seen in the self intermediate scattering function $F_s(q,t)$
(Fig.~\ref{figclustfree}b). At short times the correlator decays quickly
onto a plateau (not very pronounced) before the relaxation of the
steric hindrance starts. For $R=4$ and 6, this process ends in that the
correlator reaches the final plateau (which is given by $f_q^{\rm PN}$
discussed above). However, for $R=8$ the final decay is so slow, again
compatible with a logarithmic time dependence, that we do not see the
asymptotic behavior.

Finally we look at the motion of the non-clustered particles and compare
it with the one for $R=0$. From Fig.~\ref{figclustfree}a we recognize that
at this volume fraction also these particles are slowed down by the cage
effect in that one sees for all values of $R$ a shoulder in $\Delta^2(t)$
at time $t\approx 1$. For $R=0$ the $\Delta^2(t)$ shows then immediately
the diffusive behavior, i.e. it is proportional to $t$. However, if $R$
is increased, the $t-$dependence of $\Delta^2(t)$ for the un-clustered
particles follows first the one of the clustered particles. Only once
the latter starts to reach the plateau discussed above, do the former cross
over to the diffusive behavior. Hence we can conclude that 
before this crossover the relaxation dynamics of the two population
of particles are strongly coupled. This result is reasonable because in
order to move, the un-clustered particles have to explore the holes within
the percolating cluster formed by the particles which are permanently linked.

Also the self intermediate scattering function of the non-clustered
particles tracks the one of the clustered particles at short times
(Fig.~\ref{figclustfree}b). However, once the latter starts to show at
long times a plateau that has a significant height, the two correlators
differ strongly since the one for the non-clustered particles decays
to zero at long times. From the graph we also see that for $R=8$
the correlator is extremely stretched and shows almost a logarithmic
$t-$dependence. This very slow decay indicates that the mobile clusters
can move around the percolating cluster only with great difficulty, a
behavior that is similar to the relaxation dynamics of particles moving
in random porous media~\cite{kim,kurzidim}.  It is also interesting that
for the highest $R$ the mean squared displacement shows for the last two decades
in time a nice diffusive behavior, whereas $F_s(q,t)$ is far from having
decayed to zero. This apparent contradiction is related to the fact that
$\Delta^2(t)$ is dominated by the particles that move relatively fast
(i.e. they are in the small clusters) whereas $F_s(q,t)$ is dominated by
the slowly moving particles (i.e. the large clusters). This cluster-size
dependent dynamics leads to a so-called dynamical heterogeneity and
in Section~\ref{sec_vi} we will discuss this phenomenon in more detail.

Although we show in~Fig.~\ref{figclustfree} the comparison between the dynamics
of clustered and non-clustered particles for the case that $\tau_{\rm
link}$ is infinitely large, it is evident that for a very large but {\it
finite} value of $\tau_{\rm link}$ the relaxation dynamics will be very
similar. Hence if, e.g., $\tau_{\rm link}$ is on the order of $10^5$,
basically none of the shown curves will change significantly and thus 
the conclusions drawn from Fig.~\ref{figclustfree} will be apply also
for such value of $\tau_{\rm link}$.

\subsection{Relaxation timescales}

We now investigate how the two different relaxation timescales, one due
to breaking of local cages and the other due to the reconfiguration
of the network-bonds, vary with the volume fraction $\phi$ and the
connectivity $R$.

\begin{figure}[]
\includegraphics[width=70mm,clip=true]{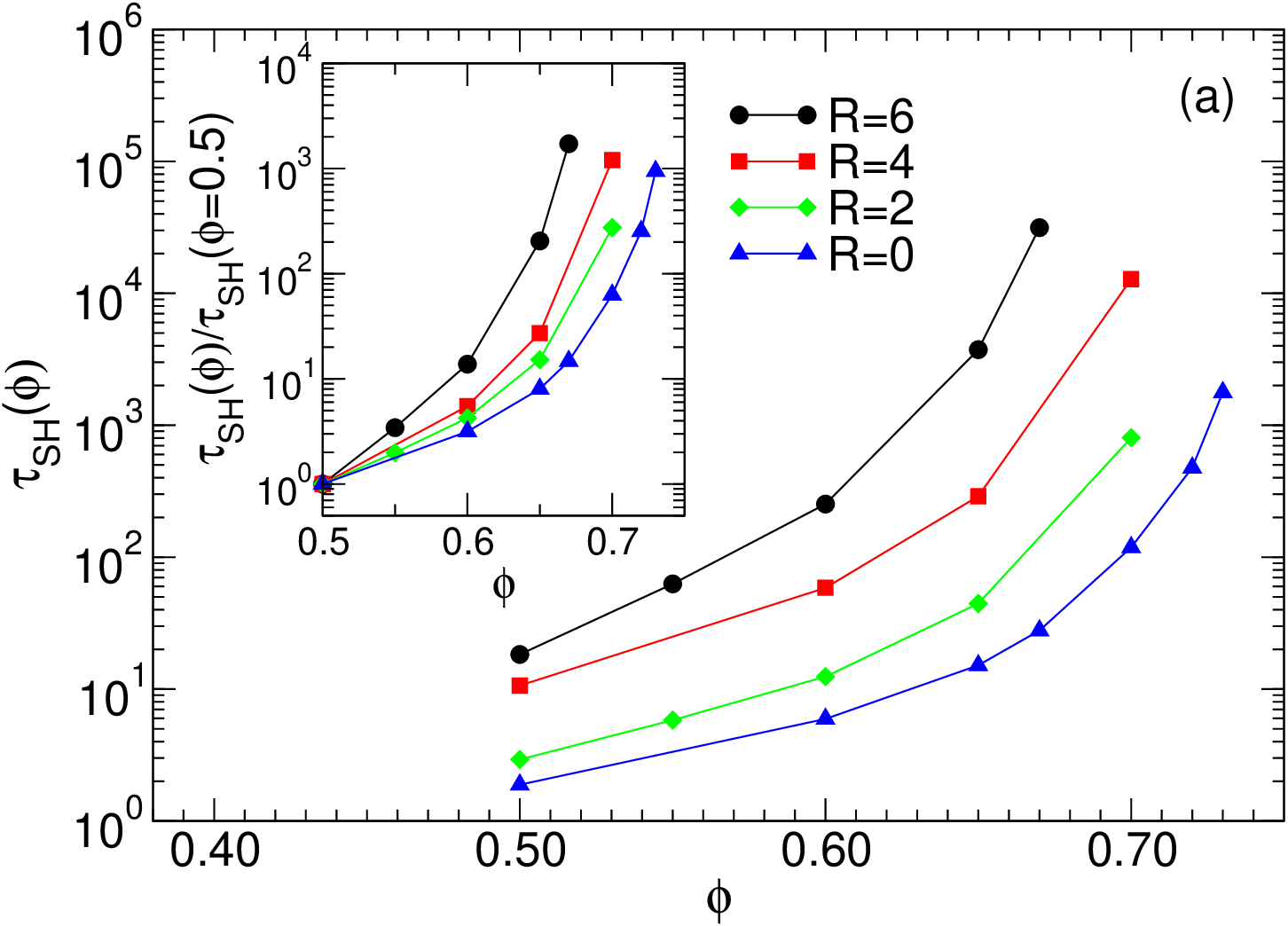}
\includegraphics[width=72mm,clip=true]{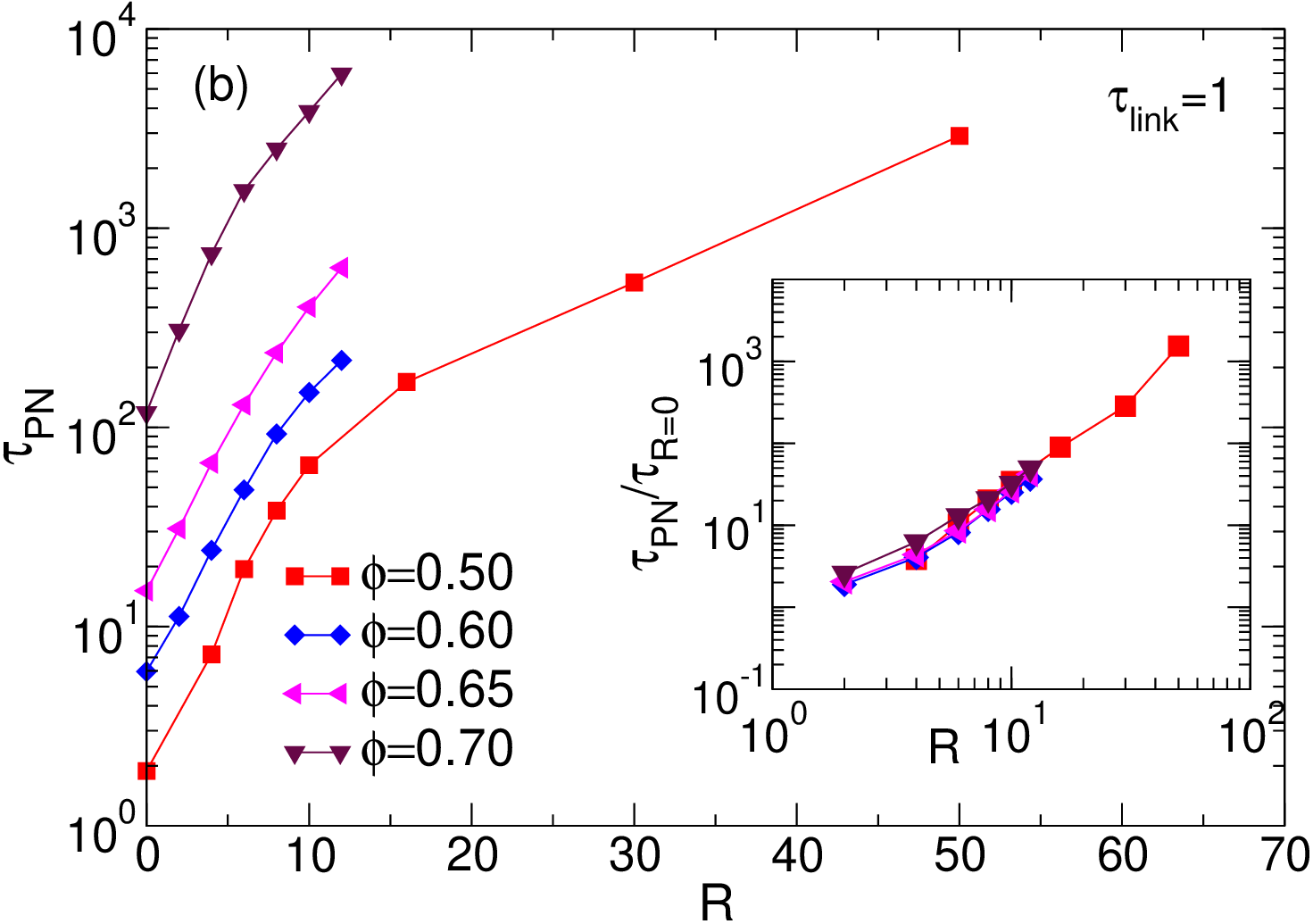}
\caption{(Color online) 
(a) Main panel: Plot of $\tau_{\rm SH}$ vs. $\phi$ for different values
of $R$. Inset: Same data scaled by $\tau_{\rm SH}(q,\phi=0.5)$.  (b)
Main panel : $\tau_{\rm PN}$ vs. $R$, for different values of $\phi$.
Inset: Same data as in the main panel but normalized by the relaxation
time for $R=0$, leading to a master curve.  Note the double logarithmic
scale in this plot.}
\label{figrelaxtime}
\end{figure}

To start we consider the case of structural relaxation related to the
steric hindrance. In order to avoid that this relaxation process is
influenced by the one of the network, we consider the case in which
the latter is completely suppressed, which can be achieved by choosing
$\tau_{\rm link}=\infty$. In the following we will study the relaxation
times associated with the intermediate scattering function for wave-vector
$q=6$.  As discussed above, this correlator shows at long times an
asymptotic plateau the height of which, $f_q^{\rm PN}$, depends on $R$ and
$\phi$. To take this into account we define the relaxation time $\tau_{\rm
SH}$, to be the time at which $F_s(q,t)-f_q^{\rm PN}=0.03$.  The evolution
of $\tau_{\rm SH}$ with $\phi$ is shown in Fig.~\ref{figrelaxtime}a,
for different value of $R$.

We see that in the absence of any bonds, i.e. for $R=0$, we have the usual
slowing down of dynamics with increasing $\phi$. The $\phi-$dependence
of the relaxation time can be fitted by a Vogel-Fulcher-Tammann-law of
the form $\tau_{\rm SH}\sim{\exp[A/(\phi_c-\phi)^\beta]}$, with $\beta
\approx 1$, from which we can estimate the volume fraction $\phi_c$ at
which the relaxation times would diverge. If we increase the number of
bonds among the particles, we see that $\tau_{\rm SH}$ increases. That
this increase is not just a constant ($R-$dependent) factor but depends
also on $\phi$ is demonstrated in the inset of the figure where we
have normalized the relaxation times to its value at $\phi=0.5$. Using
the Vogel-Fulcher-Tammann-law we can thus extract from these data the
$R-$dependence of $\phi_c$, which can be considered as a proxy for the
$R-$dependence of the glass transition temperature. This $\phi_c(R)$ line
is included in Fig.~\ref{phasediagram} as well and we see that it has a
weak negative slope with $\phi_c(R=0)=0.847$ and $\phi_c(R=6)=0.808$. Thus
we can conclude that the glass transition as induced by the steric
hindrance mechanism does not depend strongly on  the value of $R$.
However, since the prefactor of the Vogel-Fulcher-law does depend strongly
on $R$, we can conclude that a line of iso-relaxation time will bend
significantly more.

Next, we investigate how the timescale for the full relaxation of the
network of particles depends on the average number of bonds, $R$.
For this we define a relaxation time using the $t-$dependence of
$F_s(q=6,t)$ with a very short lifetime for the bonds ($\tau_{\rm
link}=1$). $\tau_{\rm PN}$ is then defined via $F_s(q=6,\tau_{\rm
PN})=0.03$ and the $R-$dependence of this relaxation time is shown
in Fig.~\ref{figrelaxtime}b. The figure shows that for small and
intermediate values of $R$ the $R$-dependence is independent of the
volume fraction in that the curves for the different values of $\phi$
seem to be just shifted vertically. That this is indeed the case is
demonstrated in the inset where we show that a plot of the same data,
but now normalized to the relaxation time for $R=0$, gives a master
curve. The main panel shows that at small concentration of bonds the
$R-$dependence is close to an exponential, a result that is likely
related to the fact that increasing $R$ leads to a tighter cage for the
steric hindrance and hence a slower dynamics.  However, for intermediate
values of $R$ the curve $\tau_{\rm PN}$ start to bend over towards a
weaker $R-$dependence. In order to investigate this effect better we
have carried out simulations for $\phi=0.50$ at very high values of $R$:
16, 30, and 50. We find that in this regime the relaxation time follows
closely an exponential (see main Fig.~\ref{figrelaxtime}b), but with an
exponential scale that is smaller than the one seen at small $R$.
Note that this dependence implies that there is no singularity in the
relaxation dynamics at any finite value of $R$, at least for this volume
fraction.  Instead the dynamics shows a behavior that is similar to an
Arrhenius law in that the barrier for the relaxation depends only on the
number of bonds between the particles. This observation is in agreement
with earlier simulations for equilibrium gels \cite{demichele}.

\begin{figure}
\includegraphics[width=80mm]{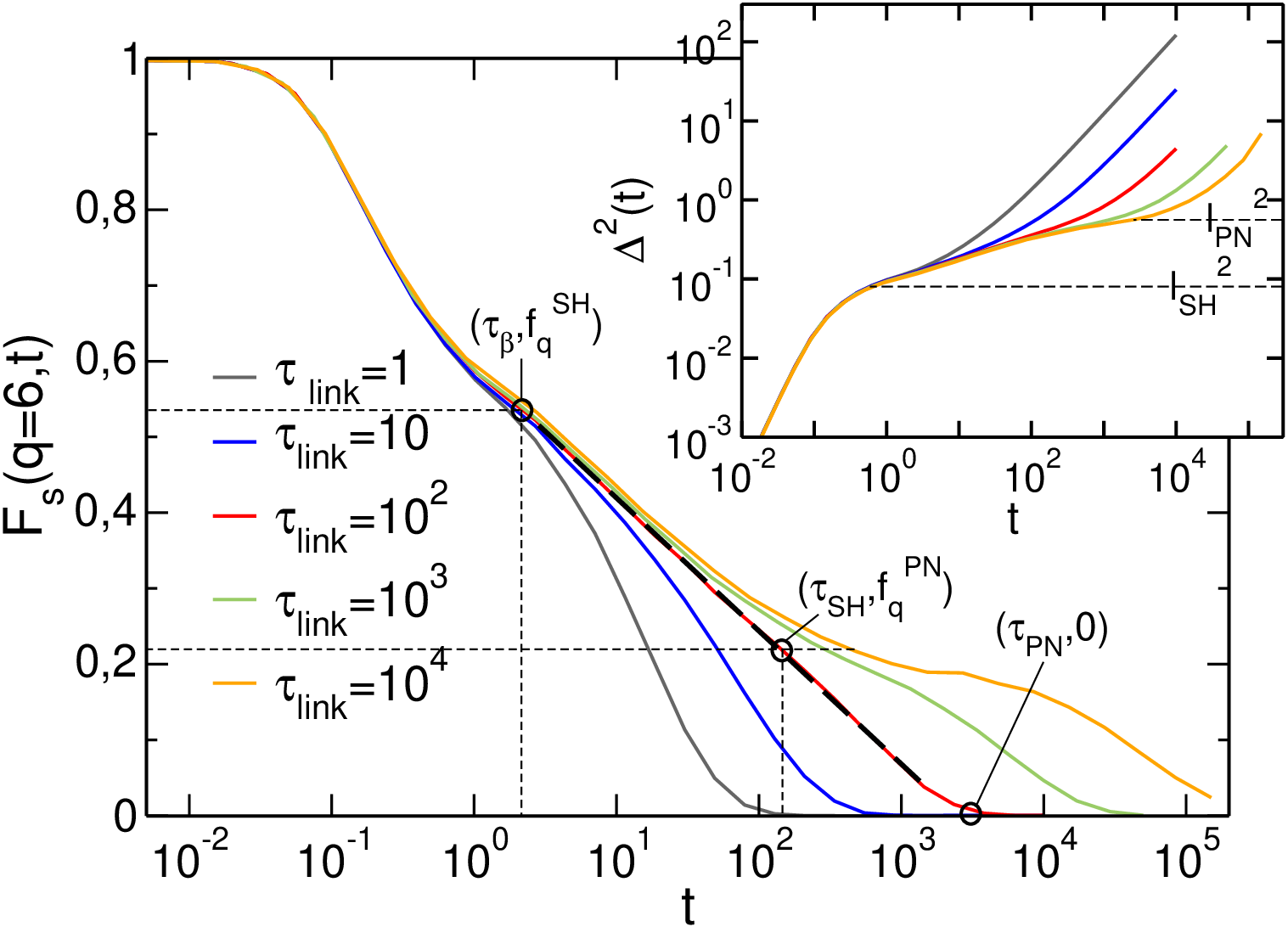}
\caption{ (Color online) \label{decon} $F_s(q,t)$ vs. $t$ for $R=6.3,
\phi=0.61$, and different $\tau_{\rm link}$. The inset shows $\Delta^2(t)$
for the same parameters.  The dotted lines in the main figure mark
the plateaus in $F_s(q,t)$: $f_q^{\rm SH}$ and $f_q^{\rm  PN}$, which
respectively correspond to the localization lengths $l_{\rm SH}$ and
$l_{\rm PN}$ indicated in the inset.}

\end{figure}

We conclude this discussion by using the different relaxation timescales
to develop a criterion that tells whether or not the correlator $F_s(q,t)$
will show the logarithmic $t-$dependence discussed in the context of
Fig.~\ref{fsq6t100}b. For this we define $f_q^{\alpha}$ as the height
of the plateau associated to the $\alpha$-process, with $\alpha\in
\{{\rm SH, PN}\}$, and recall that $\tau_{\beta}$, $\tau_{\rm SH}$
and $\tau_{\rm PN}\propto \tau_{\rm link}$ are the timescales for
the three different relaxation processes described above, namely the
rattling inside the cage, the escape from the local cage, and the network
renewal process. We can now define a simple criterion for the anomalous
logarithmic relaxation to be observed by requiring that the slope of
the two segments defined by pairs $[(\ln \tau_{\beta},f_q^{\rm SH}),(\ln
\tau_{\rm SH},f_q^{\rm PN})]$ and $[(\ln \tau_{\rm SH},f_q^{\rm PN}),(\ln
\tau_{\rm PN},0)]$ is the same, see Fig. \ref{decon}. It is known that
for glass-forming systems the plateau related to steric hindrance is a
Gaussian functions of the wave-vector $q$, $f_q^{\rm SH}\sim \exp(-q^2
\ell_{\rm SH}^2)$~\cite{nauroth}. We showed earlier, see Fig.~\ref{ap1}c,
that for the PN-process,  $f_q^{\rm PN}$ has an exponential shape at large
$q$ (the regime corresponding to anomalous logarithmic behavior). Putting
these elements together we thus obtain

\begin{equation}
 (q\ell_{\rm PN} - q^2\ell_{\rm SH}^2) = \ln \left[1 + 
\frac{\ln (\tau_{\rm SH}\tau_{\beta}^{-1})}{\ln (\tau_{\rm PN}\tau_{\rm SH}^{-1})} \right] \, .
\label{eq1}
\end{equation}

\noindent
This equation relates a purely structural observable which depends
on the competition between two localization lengthscales with
dynamical information encoded in the different relaxation times,
predicting a precise connection between structure and dynamics
whenever anomalous logarithmic relaxation is to be expected.

\section{Heterogeneities in dynamical properties}
\label{sec_vi}

Typical glass formers show a heterogeneous dynamics of the particles when
the system is increasingly supercooled. It reflects the broad distribution
in the timescales for local structural relaxations \cite{chi4ref}. On the
other hand, we have reported earlier \cite{pablo} that also the dynamics
in the gel phase can be heterogeneous: The particles in the percolating
cluster and those that are unattached have different nobilities till
the timescales at which the bonds in the network are reconfigured. In
the following we will thus explore how these two different sources of
heterogeneity interact as we increase the density of particles as well
as the number of connectivities in the system.

We use two different measures of dynamical heterogeneity.  The
first one is the non-Gaussian parameter $\alpha_2(t)$, defined as
$\alpha_2(t)=3\langle r^{4}(t) \rangle/5\langle r^{2}(t) \rangle ^{2}-1$,
where $\langle r^{2}(t) \rangle$ and $\langle r^{4}(t) \rangle$ are the
second and fourth moments of the distribution function of single particle
displacements $G_s(r,t)=N^{-1}\sum_i\langle \delta(r-|{\bf r}_i(t) -
{\bf r}_i(0)|)\rangle$, i.e. of the self part of the van Hove function.
A non-zero value of $\alpha_2(t)$ quantifies the extent of deviation
from a Gaussian shape for $G_s(r,t)$ \cite{alpha2}. Note that $G_s(r,t)$
is Gaussian at short times, i.e. in the ballistic regime, and again at
long times when the particles are diffusive.

\begin{figure}[ht]
\includegraphics[width=80mm,clip=true]{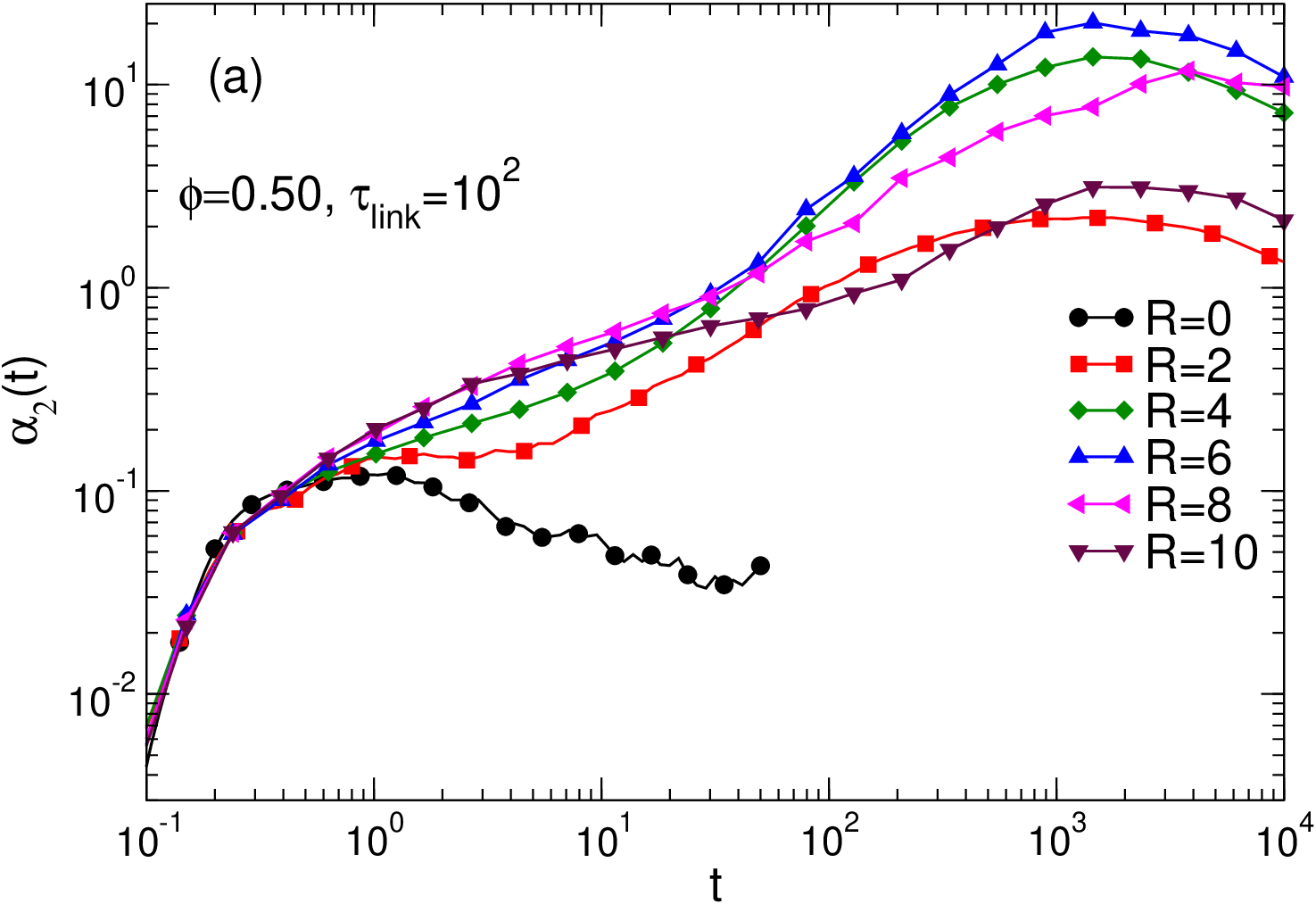}
\includegraphics[width=80mm,clip=true]{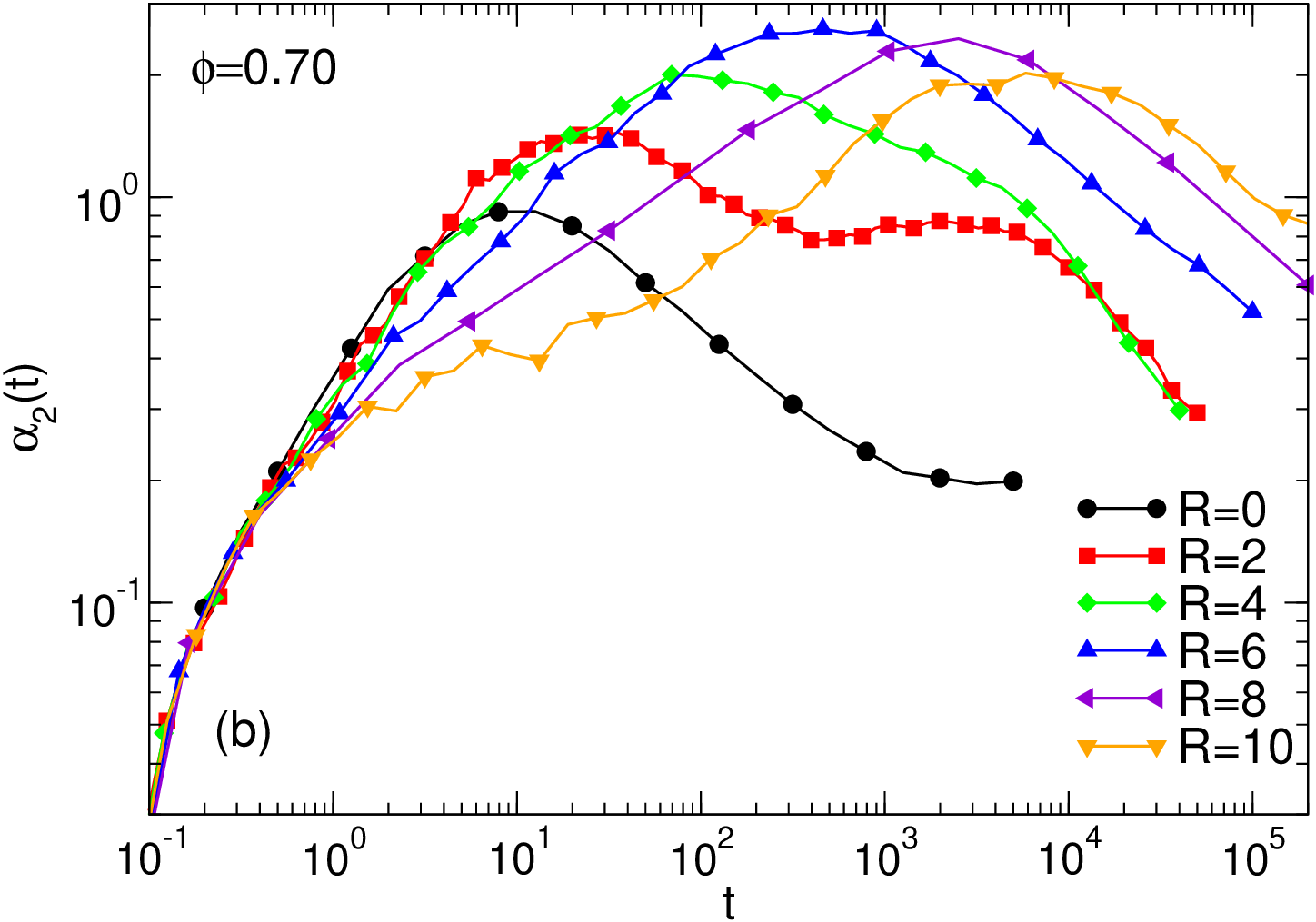}
\caption{(Color online) 
Time dependence of $\alpha_2$ for different values of $R$ (as marked
in the graphs) for $\tau_{\rm link}=10^2$. (a): $\phi=0.50$ in
double-logarithmic representation and (b): $\phi=0.70$.}
\label{alpha2phi}
\end{figure}

The second measure is the dynamic susceptibility $\chi_4(q,t)$,
computed via the fluctuations of time-correlation function:
$\chi_4(q,t)=N[\langle{F^2_s(q,t)}\rangle-\langle{F_s(q,t)}\rangle^2]$.
It is designed to capture the spatiotemporal correlations of particle
nobilities and provides, from its peak value, an estimate of the dynamic
correlations \cite{glotzer, chi4ref}.  Such functions have also been
analyzed for gels with permanent bonds \cite{coniglio3} or low density
gels \cite{colombo2}.

\subsection{Non-Gaussian parameter $\alpha_2(t)$}

We begin by looking at how the non-Gaussian parameter varies with changing
connectivities $R$, either when we go from the liquid to the gel phase, at
$\phi=0.50$, or when we are in a strongly glassy region, $\phi=0.70$. For
the case $\tau_{\rm link}=10^2$ the data is shown in Fig.~\ref{alpha2phi}.

For $\phi=0.50$, Fig.~\ref{alpha2phi}a,  $\alpha_2(t)$ shows at $t\approx
1$ a peak if $R=0$ and a shoulder for $R>0$. This feature is thus related
to the dynamical heterogeneity due to the steric hindrance mechanism,
as in usual glass-forming systems~\cite{kob_95}. For $R>0$ we find
in addition a very prominent peak the location of which is basically
independent of $R$, which shows that for this packing fraction the time
at which the system is maximally heterogeneous does not depend on $R$.
This observation is in tune with our earlier discussion that the timescale
for structural relaxation at small values of $\phi$ does not change with
$R$ if the number of bonds is not very large (see Fig.~\ref{fsq6t100}a).
The maximum occurs around $t\approx{10^3}$, a time which corresponds to
a timescale for which a significant number of particles have broken the
bonds with their neighbors and exit the constraints of the network. The
location of the peak is thus somewhat larger than $\tau_{\rm link}$. As
has been documented in Ref.~\cite{pablo}, at small packing fractions
one has on this timescale two families of particles, one for the
mobile particles and the other related to those that are immobile
and as a consequence the shape of $G_s(r,t)$ deviates strongly from a
Gaussian. Figure~\ref{alpha2phi}a indicates thus that the same behavior
persists to the larger volume fraction of $0.50$. One interesting feature
is that the peak height is non-monotonous in $R$ in that it increases
till $R=6$ and then decreases again for larger $R$ values. The reason
for the growth is that an increasing $R$ allows for more diverse values
of the connectivity for the particles, and hence to a stronger variety
in the dynamical behavior. On the other hand, if $R$ is very large most
of the particles are strongly connected all the times and hence show a
much smaller variation in their relaxation dynamics, i.e. they behave
like in a mean-field like regime.

Next we study the behavior for $\phi=0.70$, Fig.~\ref{alpha2phi}b.
When there are no bonds, $R=0$, we see that $\alpha_2(t)$ is peaked
at around $t=8$. This peak, which is related to the steric hindrance,
corresponds thus to the one seen for $\phi=0.50$ at $t\approx 1$ and
which, due to the higher density has shifted to larger times.  For $R=2$
(when the percolating cluster of the bonds develops), the location of
this peak slightly shifts to larger times and one sees the appearance
of a second peak at $t \approx 2000$ (corresponding to the timescales
for renewal of the network connectivities). However, the dominant
heterogeneity is still due to the local steric hindrances.  As $R$ is
increased, the location of the first peak continues to shift to longer
timescales, in track with the increasing structural relaxation timescales
(see Fig.~\ref{fsq6t100}c), and also its height increases, in qualitative
agreement with the behavior found in simple glass-formers if the coupling
is increased~\cite{kob_95}. For $R=6$, the two peaks have merged, since
the timescales for the two sources of heterogeneity are nearly the same,
and thus $\alpha_2(t)$ has a single peak. If $R$ is increased even more,
the position of the peak moves to larger times, but its height starts to
decrease. The reason for this decrease is likely the same as the one we
indicated when we discussed the data for $\phi=0.50$, i.e. that for large
$R$ the system starts to become mean-field like and hence heterogeneities
are suppressed.

We also note that the height of the peak is significantly smaller
than the one for $\phi=0.50$. Thus, at large density, steric hindrance
dominates and even the faster particles have less space to move around
resulting in significantly less non-Gaussian shapes for $G_s(r,t)$. As
a consequence, the dynamics is more homogeneous in this regime compared
to the gel at lower $\phi$. 

\begin{figure}[]
\includegraphics[width=80mm,clip=true]{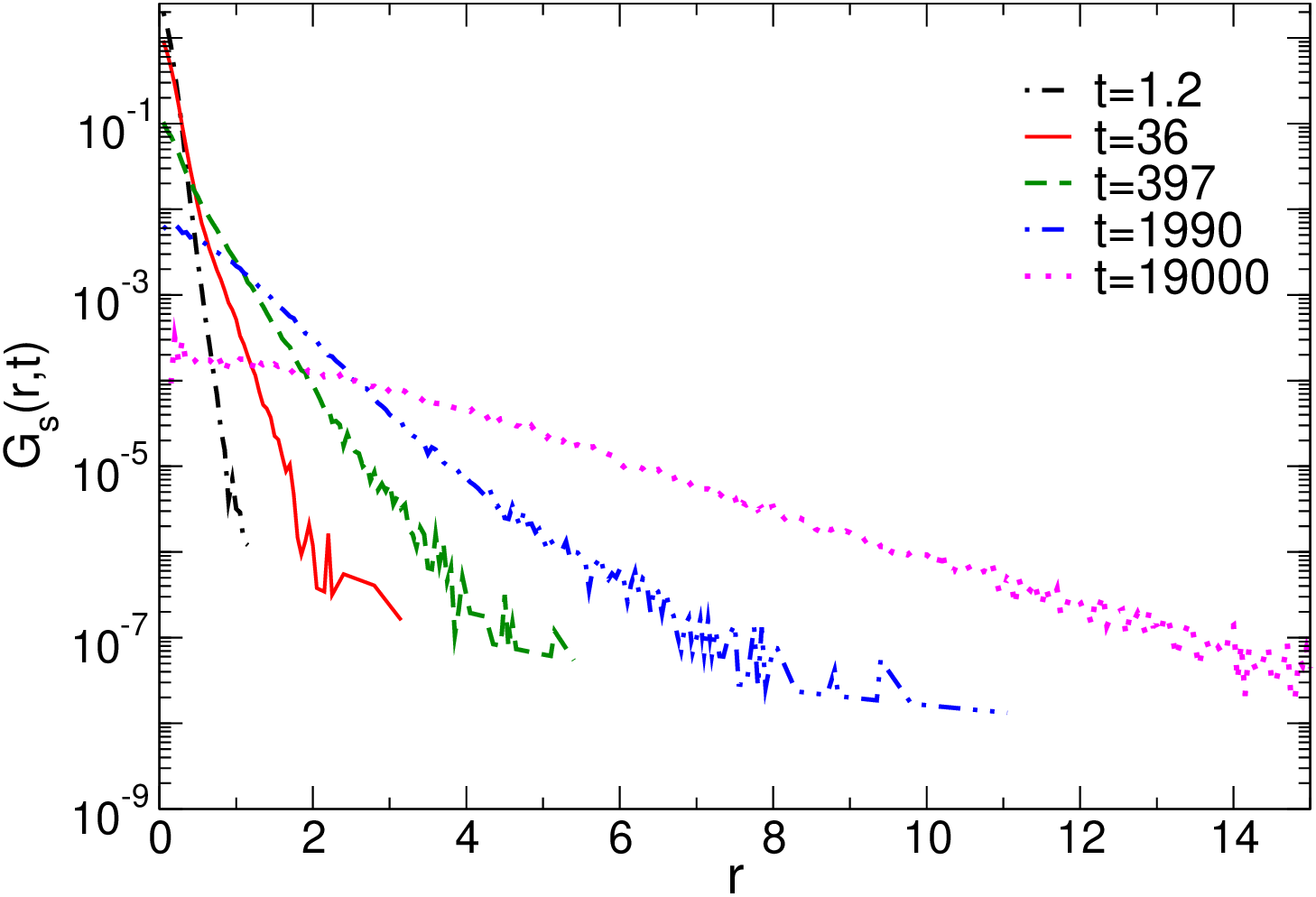}
\caption{(Color online) Self part of the van Hove function for $R=2$
and $\phi=0.70$, parameters for which $\alpha_2(t)$ shows two peaks (see
Fig.~\ref{alpha2phi}b).}
\label{vanhover4pv70}
\end{figure}

To elucidate the origin of the two peaks in the non-Gaussian
parameter at $R=2, \phi=0.70$, we determine how the distribution of
particle displacements $G_s(r,t)$ evolves with time. This is shown in
Fig.~\ref{vanhover4pv70}. At short times,  particle motion is restricted
by the local cage of neighbors and thus $G_s(r,t)$ is a Gaussian at small
$r$. At larger distances the distribution has an exponential tail which
is due to a few particles that have escaped the steric hindrance cage,
in agreement with the usual heterogeneous glassy dynamics in supercooled
systems~\cite{Pinaki}. The presence of these two processes gives rise
to the maximum in $\alpha_2$ at short times. At around $t=400$, most of
the particles have escaped from this steric hindrance cage and thus the
dynamics of the system becomes more homogeneous with $G_s(r,t)$ assuming
a Gaussian form and hence $\alpha_2$ decreases again. But with time, the
motion of the particles is again restricted, this time by the bonds which
have not yet relaxed. Thus again $G_s(r,t)$ develops a Gaussian shape
(with a larger width, determined by the length of the connecting bonds)
and an exponential tail, implying an increase in $\alpha_2$.  Later,
the network eventually relaxes, the particles are diffusive and thus
$\alpha_2$ decreases again.

\begin{figure}[]
\includegraphics[width=80mm,clip=true]{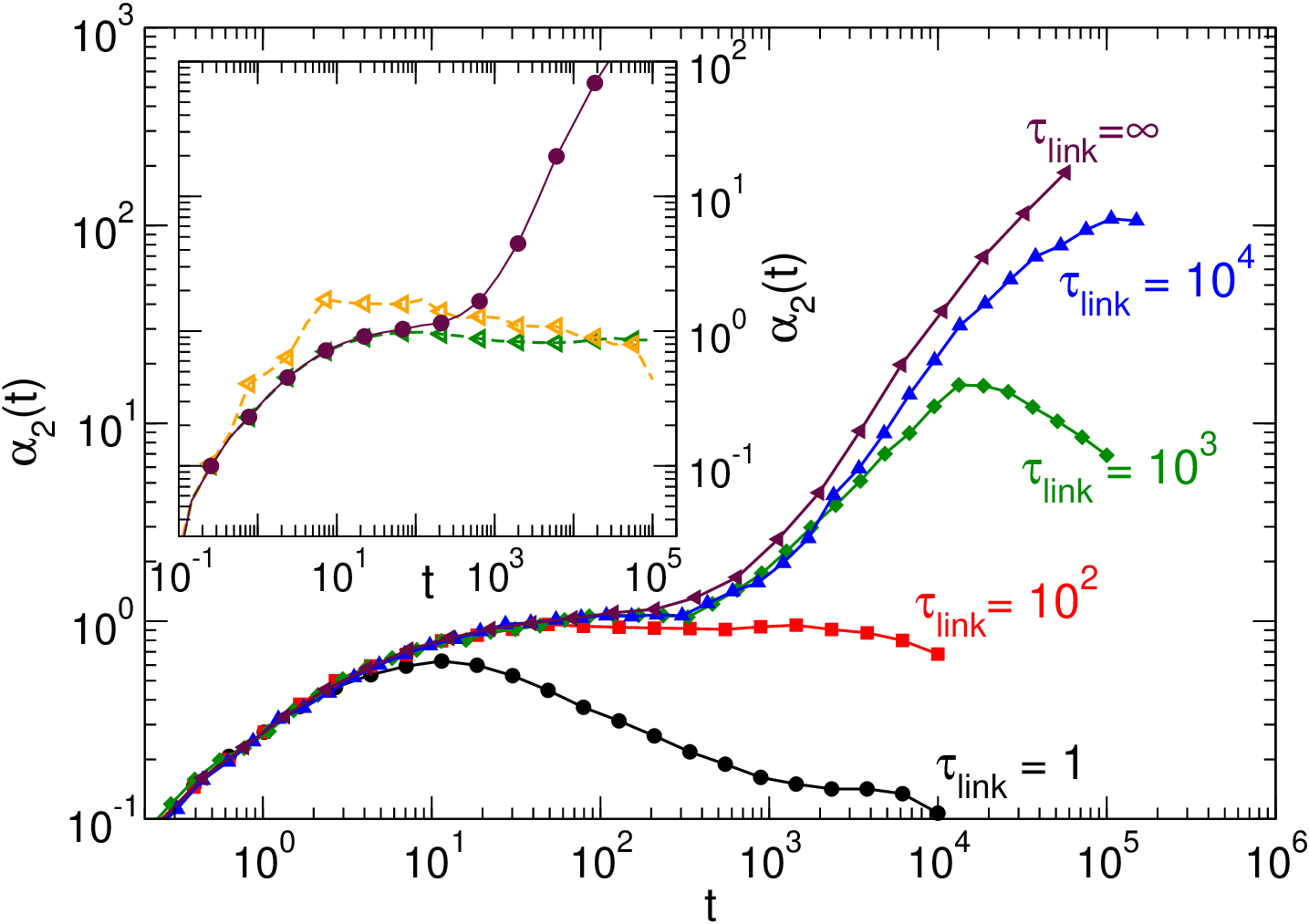}
\caption{(Color online) Time dependence of $\alpha_2$ for different
values of $\tau_{\rm link}$ at  $R=6, \phi=0.65$.  Inset:  For the case
of fixed bonds ($\tau_{\rm link}=\infty$), $\alpha_2$ for the full system
(maroon) along with those for the clustered (green open symbols) and
non-clustered particles (orange filled symbols). }
\label{alpha2taulink}
\end{figure}

The presence of these two different contributions to the shape of
$\alpha_2(t)$ can be further clarified by varying the lifetime of the
network, i.e. $\tau_{\rm link}$, and in Fig.~\ref{alpha2taulink} we
show this for the case $\phi=0.65, R=6$. Since we scan a large span
of timescales and $\alpha_2$ changes strongly, the data is shown in
logarithmic scales. For $\tau_{\rm link}=1$, we see only a small peak at
$t\approx 10$, which corresponds to the one due to the steric hindrance.
For $\tau_{\rm link}=10^2$, we find that this peak has increased a bit
in height and its location has also shifted to $t \approx 30$. This
change is a consequence of the increased effective coupling between the
particles due to the increased lifetime of the bonds. In addition, we
see the development of a new (weak) peak at $t\approx{10^3}$, which is
caused by the network renewal process.  Now, if we increase $\tau_{\rm
link}$ further, the location of the first peak remains at the same place,
since the local crowding effects are unaffected by the network dynamics.
In contrast to this, the location of the second peak shifts to longer
timescales and its height increases strongly. Since the position of
this peak scales with $\tau_{\rm link}$, we recognize that this peak
is related to the renewal process. Finally, if we take the case of
permanent bonds between the particles, the curve is nearly identical to
that for $\tau_{\rm link}=10^4$, except that we see no signature of it
eventually decreasing with time.  The reason for this runaway effects is
the fact that with increasing lifetime of the network, the unattached
particles diffuse away and travel long distances before the network is
again renewed. This results in very long extended tails in $G_s(r,t)$
which shows up as very large values for the non Gaussian parameter.
For the case that $\tau_{\rm link}$ diverges the dynamics never becomes
Gaussian and $\alpha_2(t)$ diverges at long times.

It is interesting that if one computes, for $\tau_{\rm link}=\infty$,
$\alpha_2(t)$ separately for the clustered and non-clustered particles,
only the peak at short times is observed, i.e. the one which is due to the
crowding effects. This is shown in the Inset of Fig.~\ref{alpha2taulink}
for the case of the system of particles where there is a permanently
frozen percolating cluster and few unattached particles. In this case,
the $\alpha_2(t)$ for the clustered particles follows the curve for
the full system, shows the bump at short times and becomes at long
times a constant (see the corresponding data for $\Delta^2(t)$ in
Fig.~\ref{figclustfree}). The result that at long times $\alpha_2(t)$
does not go to zero, as it would be the case for vibrations in a typical
amorphous solid with $R=0$, is related to the fact that the spanning
cluster is a disordered network of particles that have different
local environments. Thus the large variety of slow floppy motions,
related to the slow relaxation in $F_s(q,t)$ over long times (see
Fig.~\ref{figclustfree}), gives a non-Gaussian shape for $G_s(r,t)$
even at very long times. This non-Gaussianity is also the reason for
the exponential shape of $f_q^{\rm PN}(q)$, as observed earlier, see
Fig.~\ref{ap1}. For the non-clustered particles, $\alpha_2(t)$ does show
a slight decrease beyond the short-time maximum which shows that the
relaxation dynamics starts to become a bit more homogeneous. However,
it cannot be expected that the non-Gaussian parameter will go to zero
even at very long times, since the mobile clusters do have a spread in
size and hence a different diffusion constant.

\subsection{Four point susceptibility $\chi_4(q,t)$}

\begin{figure}
\includegraphics[width=80mm,clip=true]{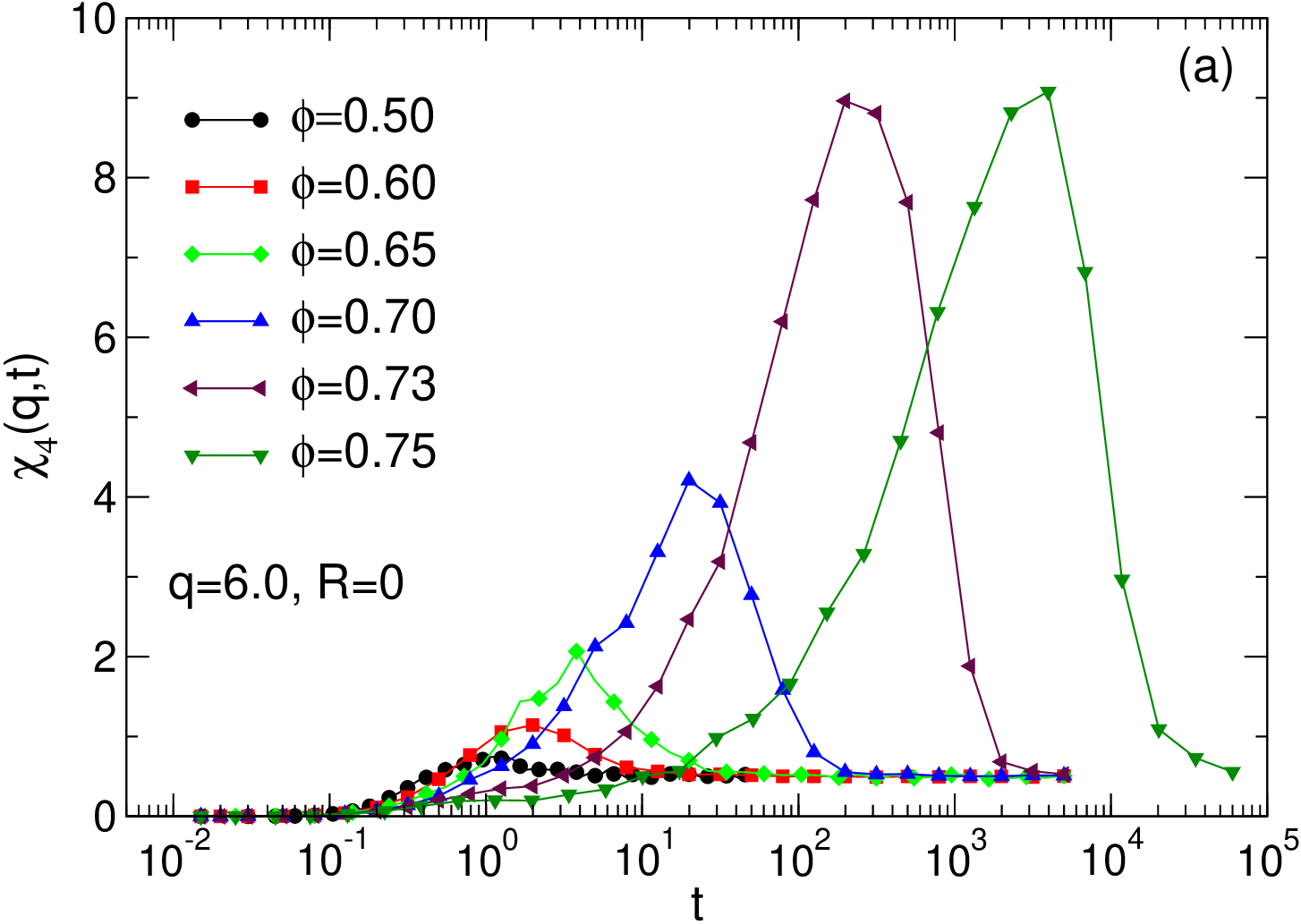}
\includegraphics[width=80mm,clip=true]{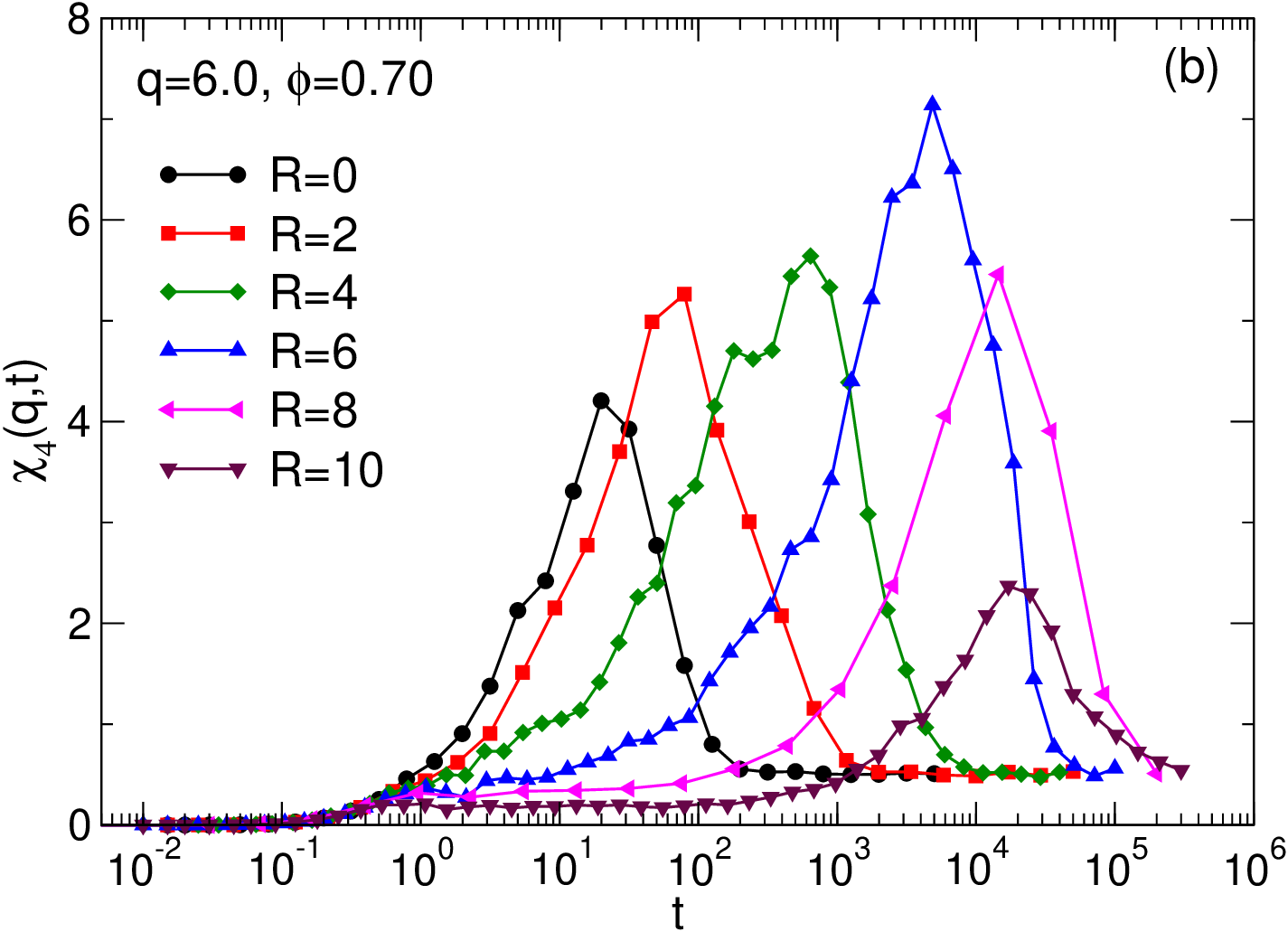}
\includegraphics[width=80mm,clip=true]{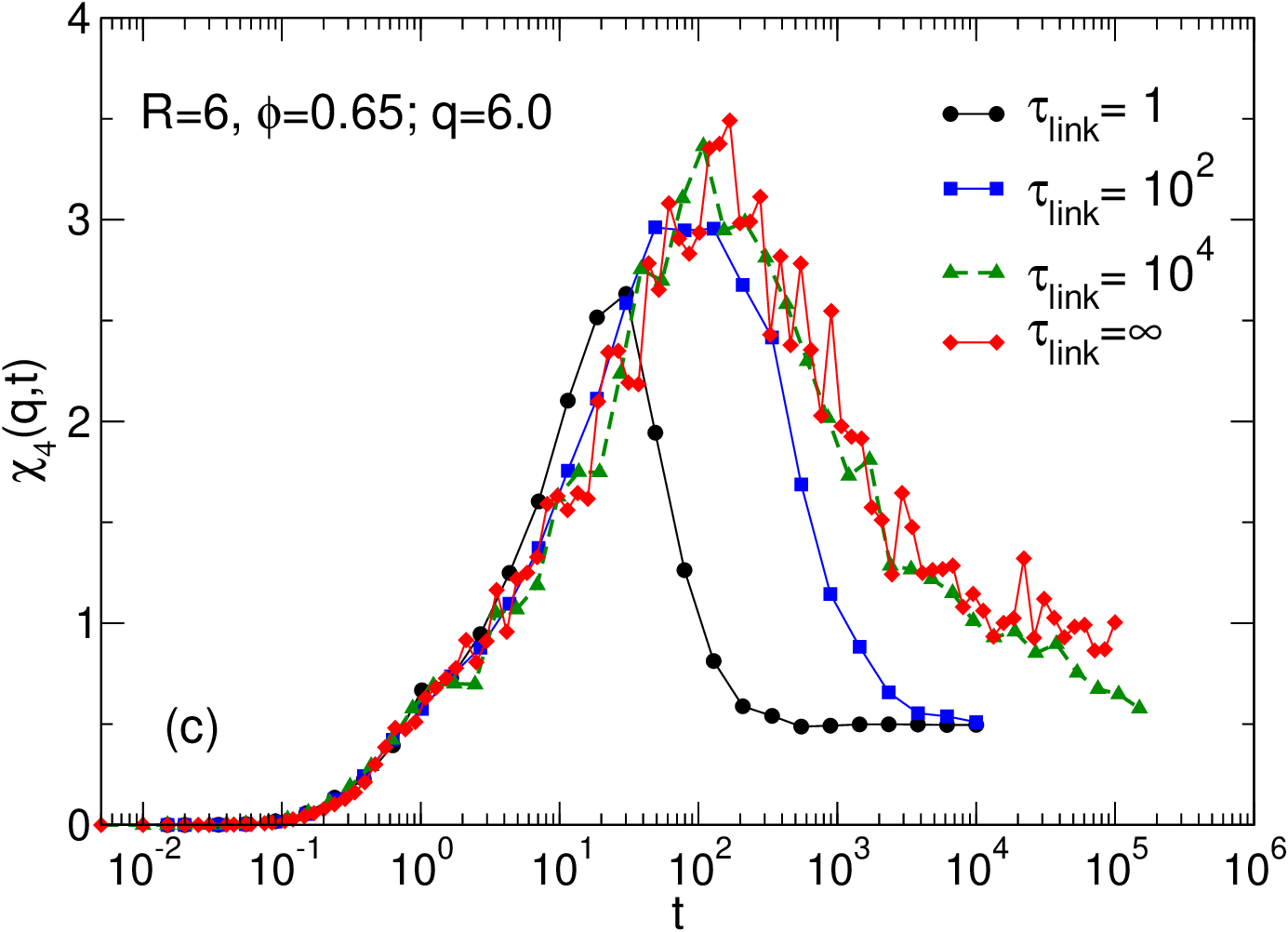}
\caption{(Color online) 
Time dependence of $\chi_4$ for $\tau_{\rm link}=100$. (a) $R=0$ and different values of $\phi$. 
(b) $\phi=0.70$ and different values of $R$ (as marked on plot) 
(c) $R=4, \phi=0.65$ for different values of $\tau_{\rm link}$ (as marked on plot).}
\label{chi4}
\end{figure}

We conclude by measuring to what extent the observed heterogeneous
dynamics is related to a correlated motion of the particles.  This can
be quantified by $\chi_4(q,t)$, defined above, measured for $q=6.0$,
i.e. we look for relaxations over distances which are slightly larger
than the average particle diameter.

To have a reference we consider first the case $R=0$, i.e. when
there are no polymers and the system is just a standard glass-forming
system. The time dependence of $\chi_4(q,t)$ is shown in Fig.~\ref{chi4}a
for different packing fractions. In agreement with earlier studies on
similar systems, Ref.~\cite{glotzer}, we find that $\chi_4(q,t)$ shows
a maximum at a time $t$ that increases with $\phi$ and which tracks the
increasing relaxation time $\tau_{\rm SH}$.  The fact that the height of
the peak increases with increasing $\phi$ indicates that the relaxation
dynamics become more cooperative, also this in agreement with previous
studies for other glass-forming systems~\cite{chi4ref}.

Next, we contrast this behavior with the case of increasing gelation,
i.e. we fix a volume fraction and increase the number of polymer
bonds. For this, we choose $\phi=0.70$, i.e. the same packing fraction
for which we have studied the correlation functions and the non-Gaussian
parameter, since for this $\phi$ the effects due to the steric hindrance
as well as the network constraints are clearly observed. The corresponding
$\chi_4(q,t)$ is shown in Fig.~\ref{chi4}b for $\tau_{\rm link}=10^2$. For
$0<R\leq 6$, the location of the peak in $\chi_4(q,t)$ shifts to larger
times and the height of the maximum increases. Thus this trend follows the
behavior observed in the data for $F_s(q,t)$ as well as $\alpha_2(t)$,
indicating that steric hindrance dominates the relaxation process and
that the dynamics becomes more and more collective. However, for $R>6$,
the peak height decreases again and its location becomes independent of
$R$. We can thus conclude that in this regime the relaxation dynamics
becomes less cooperative, a result that makes sense since the rewiring
of the network is not really a collective process. This result is also in
qualitative agreement with the decrease of the maximum in $\alpha_2(t)$,
see Fig.~\ref{alpha2phi}b, which showed that the dynamics becomes more
homogeneous.

We conclude by investigating how the lifetime of the network-bonds
influences the function $\chi_4(q,t)$. For this we have calculated this
observable for the case $\phi=0.65$ and $R=6$, i.e. at intermediate
density and in the gel phase for which one sees a significant
$\tau_{\rm link}$-dependence in the relaxation behavior (see
Fig.~\ref{fsq6t100tinf}b). We see, Fig.~\ref{chi4}c, that at small and
intermediate $\tau_{\rm link}$ the location of the peak in  $\chi_4(q,t)$
moves to larger time and that its height increases somewhat, i.e. a
behavior that is directly related to the steric hindrance mechanism in
which the effective cage becomes stiffer due to the increased lifetime
of the bonds. However, once $\tau_{\rm link}$ exceeds $10^4$, we see that
$\chi_4(q,t)$ no longer depends on this lifetime, i.e. the rewiring of
the network no longer affects the cooperativity of the dynamics and
the latter is solely dependent on the steric hindrance.

\section{Summary and Conclusions}

In this paper, we have studied a coarse-grained model for a transient
network fluid, a system that can be realized in experiments by a
oil-in-water emulsion. By tuning the volume fraction $\phi$ of the
constituent particles, the number of bonds $R$ between the mesoparticle,
and the lifetime $\tau_{\rm link}$ of these connections, we have scanned
across the phase diagram to investigate the relaxation dynamics of this
system, in particular the interplay between the gel-transition and the
glass transition induced by steric hindrance effects.

By analyzing the mean-squared displacement and self intermediate
scattering function we find that the nature of the slowing down depends
on the packing fraction: At low $\phi$ an increase of $R$ leads to a
two step relaxation with a plateau that increases continuously with $R$
and an $\alpha-$relaxation time that is independent of $R$ and that
is related to the lifetime of the bonds. In contrast to this we find at
large $\phi$ a two step relaxation with a plateau height that is basically
independent of $R$ whereas the relaxation time does depend on the density
of bonds. Thus these two different behaviors show that the system can
show a glassy dynamics that is related on one hand to a gel transition
and on the other hand to a glass transition associated with the steric
hindrance mechanism. For intermediate densities and a certain range of $R$
and $\tau_{\rm link}$ the interplay between these two mechanisms leads to
a decay of the time correlation function that is logarithmic over several
decades in time, whereas for other combinations it gives rise to a three
step relaxation. We note here that a similar multistep relaxation pattern
has been recently reported for mixtures of \emph{multiarm} telechelic
polymers and oil-in-water microemulsions \cite{malo}, suggesting the
existence of a competition of different arrest length- and time-scales
in these systems.  We have also studied how the height of the plateau at
long times, i.e. the Debye-Waller factor, depends on the wave-vector
and found that it decays in an exponential manner in $q$, i.e. very different
from the Gaussian decay observed in more standard glass-forming systems.

Furthermore we have determined the relaxation times of the system and
find that these can approximately be factorized into a function that
depends strongly on $R$ and a Vogel-Fulcher type dependence on $\phi$.
The $R-$dependent factor shows for $R\leq 10$ a strong exponential
dependence that is related to the escape of the particles from the
local cage, whereas for larger values of $R$ the dependence is weaker
and linked to an Arrhenius-like process for bond-breaking.

By studying the non-Gaussian parameter we have probed to what extent
the relaxation dynamics of the system is heterogeneous. At low packing
fraction this dynamics becomes extremely heterogeneous, if $R$ is not too
large, since some of the particles are very strongly connected to their
neighbors (and hence are immobile) whereas others can move almost freely.
However, if $R$ becomes larger than 6, the relaxation dynamics becomes
again quite homogeneous since at any time all particles are well connected
to their neighbors.  At large packing fractions the non-Gaussian parameter
shows a double peak structure and, by monitoring the van Hove function,
we can show that this feature is directly related to the two relaxation
processes, i.e. the steric hindrance and the connectivity of the network.

Whether or not the relaxation dynamics is cooperative can be characterized
by the four-point correlation function $\chi_4(q,t)$. We find that for
large packing fractions the height of the peak in $\chi_4(q,t)$ increases
rapidly with $R$, showing that the strengthened coupling leads to an
enhanced cooperative motion. However, if $R$ is increased beyond 6, this
cooperativity decreases again, since the relaxation dynamics is strongly
dominated by the rewiring of the network, i.e. a non-cooperative process.

Summarizing we see that the interplay between the two different mechanisms
giving rise to glassy behavior can lead a rather complex and unusual
relaxation dynamics. In the present study we have focused on a system in
which the particles are connected by polymers having a fixed extension
length $\ell=3.5\sigma$. For much smaller extension lengths, one can
expect to recover the re-entrant scenario observed in colloidal gels
\cite{zaccarellirev}. For longer extension lengths, the particles will
instead have more space to explore and it will likely result in a even
more floppy network. In the future, it would certainly be worthwhile to
explore also systems that have polymers with different value of $\ell$,
since this implies multiple localization lengths and thus provide novel
relaxation scenarios that can subsequently also studied in experiments.
And finally we recall that our model is motivated by an experimental
system which has highly nontrivial rheological properties, e.g. this
material flows like a liquid but eventually breaks as a brittle solid
\cite{fracture,ligoure}, being capable of self-healing through thermal
fluctuations. Thus, in the future we plan to study the rheological
properties of our model in order to understand the microscopic mechanisms
that lead to the experimental observations.

\acknowledgments
Financial support from ANR's TSANET is acknowledged. PIH acknowledges
financial support from Spanish MICINN project FIS2009-08451, FIS2013-43201-P, University
of Granada, Junta de Andaluc\'{\i}a projects P06-FQM1505, P09-FQM4682
and GENIL PYR-2014-13 project. WK is member of the Institut Universitaire
de France.

\end{document}